\documentclass[a4paper, 12pt]{article}

\usepackage{latexsym,amsmath,amsfonts,amssymb}
\usepackage{amsthm}
\usepackage{mathtools}
\usepackage{braket}
\usepackage[latin1]{inputenc}
\usepackage[american]{babel}
\usepackage{bbm}
\usepackage[nosort]{cite}
\usepackage{hyperref}
\hypersetup{linktocpage, colorlinks, citecolor=[rgb]{0,0,0.8}, linkcolor=[rgb]{0,0,0.8}, urlcolor=[rgb]{0.7,0,0.5}}

\renewcommand{\baselinestretch}{1.2}
\newcommand{\wt}{\widetilde}

\newcommand{\wb}{\overline}

\newcommand{\ds}{\displaystyle}

\newcommand{\ie}{\textit{i.e.}}

\numberwithin{equation}{section}

\newcommand{\nn}{\nonumber}

\newcommand{\be}{\begin{equation}} \newcommand{\ee}{\end{equation}}
\newcommand{\bea}{\begin{equation} \begin{aligned}} \newcommand{\eea}{\end{aligned} \end{equation}}
\newcommand{\ba}{\begin{array}} \newcommand{\ea}{\end{array}}

\newcommand{\cA}{\mathcal{A}}
\newcommand{\cB}{\mathcal{B}}
\newcommand{\cC}{\mathcal{C}}
\newcommand{\cD}{\mathcal{D}}

\newcommand{\cH}{\mathcal{H}}
\newcommand{\cI}{\mathcal{I}}
\newcommand{\cJ}{\mathcal{J}}

\newcommand{\cM}{\mathcal{M}}
\newcommand{\cN}{\mathcal{N}}

\newcommand{\cQ}{\mathcal{Q}}
\newcommand{\cR}{\mathcal{R}}

\newcommand{\cW}{\mathcal{W}}

\newcommand{\cZ}{\mathcal{Z}}

\newcommand{\bD}{\mathbb{D}}

\newcommand{\bH}{\mathbb{H}}

\newcommand{\bN}{\mathbb{N}}

\newcommand{\bQ}{\mathbb{Q}}
\newcommand{\bR}{\mathbb{R}}

\newcommand{\bT}{\mathbb{T}}

\newcommand{\bZ}{\mathbb{Z}}

\newcommand{\fsu}{\mathfrak{su}}

\DeclareMathOperator{\Tr}{Tr}

\makeatletter
\def\blfootnote{\gdef\@thefnmark{}\@footnotetext}
\makeatother

\newcommand{\numberset}[1]{\mathbb{#1}}

\newcommand{\nC}{\numberset{C}}

\newcommand{\nN}{\numberset{N}}

\newcommand{\nQ}{\numberset{Q}}
\newcommand{\nR}{\numberset{R}}

\newcommand{\nT}{\numberset{T}}

\newcommand{\nZ}{\numberset{Z}}

\newcommand{\fg}{\mathfrak{g}}
\newcommand{\fh}{\mathfrak{h}}

\newcommand{\fs}{\mathfrak{s}}

\newcommand{\fu}{\mathfrak{u}}

\newcommand{\fD}{\mathfrak{D}}

\newcommand{\fM}{\mathfrak{M}}

\newcommand{\fR}{\mathfrak{R}}

\newcommand{\bfa}{\mathbf{a}}

\newcommand{\bfi}{\mathbf{i}}
\newcommand{\bfj}{\mathbf{j}}
\newcommand{\bfk}{\mathbf{k}}

\DeclarePairedDelimiter{\abs}{\lvert}{\rvert}
\newcommand{\Abs}[1]{\left\vert#1\right\vert}
\newcommand{\RE}{\operatorname{Re}}
\newcommand{\IM}{\operatorname{Im}}

\newcommand{\rk}{\text{rk}}
\newcommand{\Res}[1]{\underset{ #1}{\displaystyle\text{Res}}}
\newcommand{\hum}{\hat{u}^{\scriptscriptstyle(\vec{m})}}

\begin{document}

\thispagestyle{empty}
\begin{flushright}
SISSA  46/2018/FISI
\end{flushright}
\vspace{10mm}
\begin{center}
{\huge  A Bethe Ansatz type formula \\[.5em] for the superconformal index}
\\[15mm]
{Francesco Benini$^{1,2,3}$, Paolo Milan$^{1,2}$}
 
\bigskip
{\it
$^1$ SISSA, Via Bonomea 265, 34136 Trieste, Italy \\[.5em]
$^2$ INFN, Sezione di Trieste, Via Valerio 2, 34127 Trieste, Italy \\[.5em]
$^3$ ICTP, Strada Costiera 11, 34151 Trieste, Italy \\[.5em]
}

{\tt fbenini@sissa.it, pmilan@sissa.it}

\bigskip
\bigskip

{\bf Abstract}\\[5mm]
{\parbox{14cm}{\hspace{5mm}

Inspired by recent work by Closset, Kim and Willett, we derive a new formula for the superconformal (or supersymmetric) index of 4d $\cN=1$ theories. Such a formula is a finite sum, over the solution set of certain transcendental equations that we dub Bethe Ansatz Equations, of a function evaluated at those solutions.
}}
\end{center}
\newpage
\pagenumbering{arabic}
\setcounter{page}{1}
\setcounter{footnote}{0}
\renewcommand{\thefootnote}{\arabic{footnote}}

{\renewcommand{\baselinestretch}{1} \parskip=0pt
\setcounter{tocdepth}{2}
\tableofcontents}

\section{Introduction and summary}

In supersymmetric quantum field theories there are many classes of observables that can be computed exactly and non-perturbatively, making supersymmetry an appealing testing ground for general ideas about quantum field theory and, through holography, also quantum gravity. One of those observables is the superconformal index \cite{Romelsberger:2005eg, Kinney:2005ej, Bhattacharya:2008zy} which---in theories with superconformal invariance---counts with signs the number of local operators in short representations of the superconformal algebra. This counting can be done keeping track of the spin and other charges of the operators. Despite its simplicity, the superconformal index is an observable that contains a lot of information about the theory, and indeed it has been studied in all possible dimensions (\ie{} up to six) and under so many angles (for reviews see \cite{Pestun:2016zxk}). In this note we focus on the four-dimensional superconformal index.

Because the index does not depend on continuous deformations of the theory, and a suitable supersymmetric generalization thereof does not depend on the RG flow, it follows that in theories that are part of a conformal manifold and have a weakly-coupled point on it, and in theories that are asymptotically free, the evaluation of the index can be reduced to a weak coupling computation.%
\footnote{There is a small caveat: the IR superconformal R-symmetry must be visible in the UV, \ie{} it should not be accidental.}
This amounts to counting all possible local operators in short representations one can write down, and then restricting to the gauge-invariant ones. In the language of radial quantization, one counts all multi-particle states in short representations on the sphere, and then imposes Gauss law. In the case of the 4d $\cN=1$ superconformal (or supersymmetric) index of a gauge theory with gauge group $G$ and matter chiral multiplets in representation $\fR$, the counting is captured by the standard formula \cite{Romelsberger:2005eg, Kinney:2005ej, Dolan:2008qi}:
\be
\label{standard formula}
\cI(p,q;v)=\frac{(p;p)_\infty^{\text{rk}(G)}(q;q)_\infty^{\text{rk}(G)}}{\abs{\cW_G}}\oint_{\nT^{\text{rk}(G)}}\frac{ \prod_{\rho_a\in\fR}\Gamma\bigl((p q)^{r_a/2}z^{\rho_a}v^{\omega_a};p,q\bigr)}{\prod_{\alpha\in\Delta}\Gamma(z^{\alpha};p,q)}\prod_{i=1}^{\text{rk}(G)}\frac{dz_i}{2\pi i z_i} \;.
\ee
Here, briefly, $p,q$ are the (complex) fugacities associated to the angular momentum, $v$ collectively indicates the fugacities for flavor symmetries, $z$ indicates the fugacities for the gauge symmetry, $r_a$ are the R-charges, and $\Gamma$ is the elliptic gamma function. All the details will be reviewed in Section~\ref{sec: 4d index review}.

In this note, inspired by recent work of Closset, Kim and Willett \cite{Closset:2017bse, Closset:2018ghr}, we show that when the fugacities for the angular momentum satisfy
\be
\label{condition on fugacities}
q^a = p^b
\ee
for some coprime positive integers $a,b$, then one can derive an alternative, very different formula for the 4d superconformal index. The condition (\ref{condition on fugacities}) can be rewritten as
\be
p = h^a \;,\qquad q = h^b
\ee
for some fugacity $h$ and coprime $a,b \in\bN$.

The new formula, that we will explain in great detail in Section~\ref{sec: BA formula}, is a finite sum over the solution set $\fM_\text{BAE}$ to certain transcendental equations---that we dub Bethe Ansatz equations (BAEs)---of a function, closely related to the integrand in (\ref{standard formula}), evaluated at those solutions. Very schematically, we prove that
\be
\label{new BA formula}
\cI(p,q;v)= \frac{(p;p)_\infty^{\text{rk}(G)}(q;q)_\infty^{\text{rk}(G)}}{\abs{\cW_G}} \sum_{z \,\in\, \fM_\text{BAE}} \sum_{\{m_i\}=1}^{ab}\cZ \bigl( z/ h^m, p,q, v \bigr) \; H( z, p, q, v)^{-1} \;.
\ee
Here the function $\cZ$ is the integrand in the standard formula (\ref{standard formula}); $\fM_\text{BAE}$ is the set of solutions, on a torus of exponentiated modular parameter $h$, to the BAEs which take the schematic form
\be
Q_i(z, p,q,v) = 1 \qquad\qquad \text{for } i=1, \dots, \rk(G)
\ee
in terms of functions $Q_i$ defined in (\ref{eq:BAEs}); the function $H$ is a ``Jacobian''
\be
H(z, p, q, v) = \underset{ij}{\det} \, \frac{\partial Q_i(z, p, q, v)}{\partial \log z_j} \;.
\ee
The precise expressions (in which we use chemical potentials instead of fugacities, in order to deal with single-valued functions) can be found at the beginning of Section~\ref{sec: BA formula}. A special case of this formula when $p=q$, namely $a=b=1$, was derived in \cite{Closset:2017bse}.

The condition (\ref{condition on fugacities}) limits the applicability of the Bethe Ansatz (BA) formula (\ref{new BA formula}) in the space of complex fugacities. Yet, as we discuss in Section~\ref{sec: continuation}, the domain of the formula is rich enough to uniquely fix the index as a continuous function (with poles) of general fugacities. We offer two arguments, one that uses holomorphy of the index and one that just uses continuity. Roughly, the reason is that the set of pairs $(p,q)$ satisfying (\ref{condition on fugacities}) is dense in the space of general complex fugacities (see Appendix~\ref{sec: dense}).

In a separate publication \cite{Benini:2018ywd} we will use the BA formula (\ref{new BA formula}) to address the large $N$ limit of the index of a specific theory, namely $\cN=4$ Super-Yang-Mills, finding some differences with previous literature. We will connect the large $N$ limit of the index to the entropy of BPS black holes in AdS$_5$.%
\footnote{The two recent papers \cite{Cabo-Bizet:2018ehj, Choi:2018hmj} also investigate the entropy of BPS black holes in AdS$_5$.}
In this way, we will extend the success of the counting of microstates of dyonic black hole in AdS$_4$ \cite{Benini:2015noa, Benini:2015eyy, Benini:2016hjo, Benini:2016rke} to the case of electric rotating black holes in AdS$_5$ \cite{Gutowski:2004ez, Gutowski:2004yv, Chong:2005hr, Kunduri:2006ek}. More generally, the new BA formula is much easier to deal with, compared to the standard integral formula, when performing numerical computations. We thus hope that it will be useful in a wider context.

The BA formula (\ref{new BA formula}) can be thought of, in some sense, as the ``Higgs branch localization'' partner of the standard ``Coulomb branch localization'' integral formula (\ref{standard formula}), using the terminology of \cite{Benini:2012ui, Benini:2013yva}. More precisely, the existence of a formula as (\ref{new BA formula}) can be justified along the lines of \cite{Benini:2015noa, Benini:2016hjo, Closset:2016arn, Closset:2017zgf, Closset:2017bse, Closset:2018ghr}. The superconformal index can be defined as the partition function of the Euclidean theory on $S^3 \times S^1$, with suitable flat connections along $S^1$ and a suitable complex structure that depends on $p,q$, and with the Casimir energy \cite{Assel:2015nca, Bobev:2015kza} stripped off.%
\footnote{Notice that the superconformal index, up to a change of variables reviewed in Section~\ref{sec: continuation}, is a single-valued function of the fugacities, while the partition function is not \cite{Bobev:2015kza}.}
The standard localization computation of the partition function leads to (\ref{standard formula}). However, when $p,q$ satisfy (\ref{condition on fugacities}) the geometry is also a Seifert torus fibration over $S^2$. Along the lines of \cite{Closset:2018ghr}, one expects to be able to reduce to the computation of a correlator in an $A$-twisted theory on $S^2$ \cite{Nekrasov:2014xaa}, which should give an expression as in (\ref{new BA formula}). In any case, we have derived the BA formula (\ref{new BA formula}) by standard manipulations of the integral expression and thus we do not rely on any such putative 2d reduction.

The note is organized as follows. In Section~\ref{sec: 4d index review} we review the standard formula for the 4d superconformal index, carefully stressing its regime of applicability. In Section~\ref{sec: BA formula} we present our new BA formula in great detail, and then we derive it in Section~\ref{sec: proof}.

\section{The 4d superconformal index}
\label{sec: 4d index review}

In order to fix our notation, let us review the standard formulation of the superconformal index \cite{Romelsberger:2005eg, Kinney:2005ej}, which counts local operators in short representations of the 4d $\cN=1$ superconformal algebra (SCA) $\fsu(2,2|1)$. Going to radial quantization, this is the same as counting (with sign) $\frac14$-BPS states of the theory on $S^3$.

The bosonic part of the superconformal algebra is $\fsu(2,2) \oplus \fu(1)_R$, where the first factor is the 4d conformal algebra and the second one is the R-symmetry. We pick on $S^3$ one Poincar\'e supercharge, specifically $\cQ = \wb Q_-$, and its conjugate conformal supercharge \mbox{$\cQ^\dag = S_+$}. Together with $\Delta = \frac12 \{\cQ, \cQ^\dag\}$ they form an $\fsu(1|1)$ superalgebra. The superconformal index is then equal to the Witten index
\be
\label{eq:Wittenindex}
\cI(t) = \Tr_{\cH[S^3]}\,(-1)^F\,e^{-\beta\Delta} \prod\nolimits_k t_k^{J_k} \;,
\ee
where $J_k$ are Cartan generators of the commutant of $\fsu(1|1)$ in the full SCA and $t_k$ are the associated complex fugacities. By standard arguments \cite{Witten:1982df}, $\cI(t)$ counts only states with $\Delta=0$, \ie~annihilated by both $\cQ$ and $\cQ^\dagger$, and thus it does not depend on $\beta$. On the other hand, it is holomorphic in the fugacities $t_k$, which serve both as regulators and as refinement parameters.

To be more precise, the states counted by \eqref{eq:Wittenindex} have $\Delta=E-2j_+-\frac{3}{2}r=0$, where $E$ is the conformal Hamiltonian or dimension, $j_\pm$ are the Cartan generators of the angular momentum $\fsu(2)_+\oplus\fsu(2)_-\subset \fsu(2,2)$, and $r$ is the superconformal $U(1)_R$ charge. Moreover, the subalgebra of $\fsu(2,2\vert1)$ which commutes with $\fsu(1|1)$ has Cartan generators $E+j_+$ and $j_-$. Therefore we write
\be
\label{eq:traceindex}
\cI(p,q)=\Tr_{\Delta=0}\,(-1)^F\,p^{\frac{1}{3}(E+j_+)+j_-}q^{\frac{1}{3}(E+j_+)-j_-}=\Tr_{\Delta=0}\,(-1)^{F}\,p^{j_1+\frac{r}{2}}q^{j_2+\frac{r}{2}} \;,
\ee
where $j_{1,2} = j_+\pm j_-$ parametrize the rotated frame $\fu(1)_1\oplus \fu(1)_2\subset\fsu(2)_+\oplus\fsu(2)_-$ and $p,\,q$ are the associated fugacities (up to a shift by $r/2$). Whenever the theory enjoys flavor symmetries, one can introduce fugacities $v_\alpha$ for the Cartan generators of the flavor group. Then, the index will depend holomorphically also on $v_\alpha$.

The trace formula \eqref{eq:traceindex} can be exactly evaluated at all regimes in the couplings. Indeed, since $\cI$ is invariant under any continuous deformation of the theory, one can explicitly account for the contribution of every gauge-invariant state with $\Delta=0$ in the free regime \cite{Sundborg:1999ue, Aharony:2003sx, Kinney:2005ej}. In particular, the contributions of all the multi-particle states are simply encoded in the plethystic exponential \cite{Benvenuti:2006qr} of the ``single-letter partition functions'', whereas the restriction to the gauge-invariant sector is done by integrating the latter contributions over the gauge group. This procedure yields a finite-dimensional integral formula for the superconformal index, which can be expressed as an elliptic hypergeometric integral \cite{Dolan:2008qi}.%
\footnote{An alternative way to obtain the integral formula is to use supersymmetric localization \cite{Pestun:2007rz}. Indeed, the supersymmetric partition function $Z$ of the theory on a primary Hopf surface $\cH_{p,q}\simeq S^1\times S^3$ can be computed with localization \cite{Closset:2013sxa, Assel:2014paa} and it is related to the superconformal index through $Z=e^{-E_\text{SUSY}}\cI$, where $E_\text{SUSY}$ is the supersymmetric Casimir energy \cite{Assel:2015nca, Bobev:2015kza}.}

For concreteness, we consider a generic $\cN=1$ gauge theory with semi-simple gauge group $G$, flavor symmetry group $G_F$ and non-anomalous $U(1)_R$ R-symmetry. We assume that the theory flows in the IR to a non-trivial fixed point and we parametrize $U(1)_R$ with the superconformal R-charge sitting in the SCA of the IR CFT (assuming this is visible in the UV). Furthermore, the matter content consists of $n_\chi$ chiral multiplets $\Phi_a$ in representations $\fR_a$ of $G$, carrying flavor weights $\omega_a$ in some representations $\fR_F$ of $G_F$ and with superconformal R-charges $r_a$. Additionally, we turn on flavor fugacities $v_\alpha$, with $\alpha=1,\dots,\rk(G_F)$, parametrizing the maximal torus of $G_F$. The integral representation of the superconformal index is given by
\be
\label{eq:SCI}
\cI(p,q;v)=\frac{(p;p)_\infty^{\text{rk}(G)}(q;q)_\infty^{\text{rk}(G)}}{\abs{\cW_G}}\oint_{\nT^{\text{rk}(G)}}\frac{\prod_{a=1}^{n_\chi}\prod_{\rho_a\in\fR_a}\Gamma\bigl((p q)^{r_a/2}z^{\rho_a}v^{\omega_a};p,q\bigr)}{\prod_{\alpha\in\Delta}\Gamma(z^{\alpha};p,q)}\prod_{i=1}^{\text{rk}(G)}\frac{dz_i}{2\pi i z_i} \;.
\ee
The integration variables $z_i$ parametrize the maximal torus of $G$, and the integration contour is the product of $\rk(G)$ unit circles. Then $\rho_a$ are the weights of the representation $\fR_a$, $\alpha$ parametrizes the roots of $G$ and $\abs{\cW_G}$ is the order of the Weyl group. Moreover, we have introduced the notation $z^{\rho_a} = \prod_{i=1}^{\rk(G)}z_i^{\rho_a^i}$ and $v^{\omega_a} = \prod_{\alpha=1}^{\rk(G_F)}v_\alpha^{\omega^\alpha_a}$, whereas
\be
\label{eq:egamma}
\Gamma(z; p ,q) = \prod_{m,n=0}^\infty \frac{1-p^{m+1}q^{n+1}/z}{1-p^mq^nz} \;,\qquad\qquad \abs{p}<1 \;,\quad\abs{q}<1
\ee
is the elliptic gamma function \cite{math/9907061} and
\be
\label{eq:qpoch}
(z;q)_\infty = \prod_{n=0}^\infty\left(1-z q^n\right) \;,\qquad\qquad \abs{q}<1
\ee
is the $q$-Pochhammer symbol.

This representation makes manifest the holomorphic dependence of the index on $p,\,q,\,v_\alpha$. It is important to stress that the expression \eqref{eq:SCI}, which is a contour integral along $\rk(G)$ unit circles, is only valid as long as the fugacities stay within the following 
\be
\label{eq:domaininv}
\text{Domain:} \qquad\qquad \abs{p},\abs{q}<1 \;,\qquad\quad\abs{pq}<\bigl\vert  (pq)^{r_a/2}v^{\omega_a}\bigr\vert<1 \;,\qquad\quad \forall a \;.
\ee
These conditions descend from the requirement of convergence of the plethystic representation of the index, from which \eqref{eq:SCI} is derived. The plethystic expansion of the elliptic gamma function,
\be
\Gamma(z;p,q)=\exp\Biggl[\sum_{m=1}^\infty\frac{1}{m}\frac{z^m-(pq)^m z^{-m}}{(1-p^m)(1-q^m)}\Biggr] \;,
\ee
converges for
\be
\label{eq:domaingamma}
\abs{pq} < \abs{z} < 1 \qquad\text{and}\qquad  \abs{p},\abs{q}<1 \;.
\ee
The domain \eqref{eq:domaininv} then follows from requiring the integrand of \eqref{eq:SCI} to have a convergent expansion. Indeed, within the domain of convergence \eqref{eq:domaingamma}, the elliptic gamma function is a single-valued analytic function with no zeros, poles nor branch cuts.
Both $\Gamma(z;p,q)$ and $(z;q)_\infty$ can be analytically continued to $z\in\nC$. However, when we analytically continue the integral \eqref{eq:SCI} outside the domain \eqref{eq:domaininv}, the integration contour must be continuously deformed in order to take into account the movement of the various poles of the integrand in the complex plane, in such a way that the poles do not cross the contour. As a result, for generic fugacities the integration contour is not as simple as a product of unit circles. To avoid this complication, throughout this paper we will always work within \eqref{eq:domaininv}---and perform analytic continuation only at the end, if needed.

It will be useful to set some new notation. We define a set of chemical potentials through
\be
\label{eq:chempot}
p = e^{2\pi i \tau} \;,\qquad\qquad q = e^{2\pi i \sigma} \;, \qquad\qquad v_\alpha =e^{2\pi i \xi_\alpha} \;,\qquad\qquad z_i = e^{2\pi i u_i} \;,
\ee
as well as a fictitious chemical potential $\nu_R$ for the R-symmetry, whose value is fixed to
\be
\nu_R=\frac{1}{2}(\tau+\sigma)
\ee
by supersymmetry. Moreover, we redefine the elliptic gamma function as a (periodic) function of the chemical potentials:
\be
\label{eq:gammatilde}
\widetilde{\Gamma}(u,\tau,\sigma) = \Gamma\bigl(e^{2\pi i u};e^{2\pi i \tau},e^{2\pi i \sigma}\bigr) \;,
\ee
so that the integrand of \eqref{eq:SCI} can be expressed as
\be
\label{eq:Z}
\cZ(u;\xi,\nu_R,\tau,\sigma) = \frac{\prod_{a=1}^{n_\chi}\prod_{\rho_a\in\fR_a} \widetilde{\Gamma} \bigl( \rho_a(u)+\omega_a(\xi)+r_a\nu_R;\tau,\sigma \bigr)}{\prod_{\alpha\in\Delta} \widetilde{\Gamma} \bigl( \alpha(u);\tau,\sigma \bigr)} \;.
\ee
At last, we define
\be
\kappa_{G} = \frac{(p;p)_\infty^{\text{rk}(G)}(q;q)_\infty^{\text{rk}(G)}}{\abs{\cW_G}} \;.
\ee
The integral representation of the index takes then the following compact form:
\bea
\label{eq:SCIu}
\cI(p,q;v) = \kappa_{G} \int_{\bT^{\rk(G)}} \cZ(u;\xi,\nu_R,\tau,\sigma)\; d^{\text{rk}(G)}u \;.
\eea
The integration contour $\bT^{\rk(G)}$ is represented on the $u$-plane by a product of straight segments of length one on the real axes. In terms of the chemical potentials, the domain \eqref{eq:domaininv} can be rewritten as:
\be
\label{eq:domainchempot}
\IM\tau,\,\IM\sigma>0 \;,\qquad 0<\IM\omega_a(\xi)<\IM(\tau+\sigma) \;,\qquad\quad \forall a \;.
\ee
The integral formula (\ref{eq:SCIu}) is the starting point of our analysis. In the next Section we will focus our attention to the case where $\tau/\sigma$ is a rational number to derive---from \eqref{eq:SCIu}---a new formula that expresses the index as a finite sum.

\section{A new Bethe Ansatz type formula}
\label{sec: BA formula}

The integral representation \eqref{eq:SCIu} of the superconformal index is valid for generic complex values of the chemical potentials within the domain \eqref{eq:domainchempot}. However, if we restrict to a case where
\be
\label{condition rational ratio}
\tau/\sigma\in\nQ_+ \;,
\ee
we can prove an alternative formula describing the index as a finite sum over the set of solutions to certain transcendental equations, which we call {\itshape Bethe Ansatz Equations} (BAEs). We will first present the formula in detail, and then provide a proof. In Section~\ref{sec: continuation} we will also discuss the properties of the set of pairs $(\tau,\sigma)$ satisfying (\ref{condition rational ratio}).
 
Let us take
\be
\tau=a \omega \;,\qquad \sigma=b \omega \qquad\text{ with $a,b\in\bN$ such that $\text{gcd}(a,b)=1$}
\ee
and $\IM\omega>0$. This implies (\ref{condition on fugacities}). We can set $p=h^a$ and $q=h^b$ with $h = e^{2\pi i\omega}$, although we will mostly work with chemical potentials. We introduce the BAEs as the set of equations
\be
\label{eq:BAEs}
Q_i(u;\xi,\nu_R,\omega)=1 \;,\qquad\quad\forall\,i=1,\dots,\rk(G) \;,
\ee
written in terms of ``BA operators'' defined as
\be
\label{eq:Q_i}
Q_i(u;\xi,\nu_R,\omega) = \prod_{a=1}^{n_\chi} \, \prod_{\rho_a \in\, \fR_a} P\big( \rho_a(u)+\omega_a(\xi)+r_a\nu_R;\omega \big)^{\rho_a^i} \;.
\ee
The basic BA operator is
\be
\label{eq:basicBAEop}
P(u;\omega) = \frac{\ds e^{-\pi i\frac{ u^2}{\omega}+\pi i u}}{\theta_0(u;\omega)} \;,
\ee
where $\theta_0(u;\omega) = (z;h)_\infty(z^{-1}h;h)_\infty$ with $z=e^{2\pi i u}$ and $h=e^{2\pi i\omega}$. 

The BA operators satisfy three important properties. First, they are doubly-periodic in the gauge chemical potentials:
\be
\label{eq:Q_iperiodicity}
Q_i ( u+n+m\omega;\xi,\nu_R,\omega ) = Q_i(u;\xi,\nu_R,\omega) \;,\qquad\forall\, n_i,\, m_i\in\nZ \;,\quad i=1,\dots\text{rk}(G) \;.
\ee
Second, they are invariant under $SL(2,\bZ)$ modular transformations of $\omega$:
\be
\label{eq:Q_i modularity}
Q_i(u; \xi, \nu_R, \omega) = Q_i(u; \xi, \nu_R, \omega+1) =  Q_i\left( \frac u\omega; \frac\xi\omega , \frac{\nu_R}\omega, - \frac1\omega \right) = Q_i(-u; -\xi, -\nu_R, \omega) \;.
\ee
The last equality represents invariance under the center of $SL(2,\bZ)$. Third, they capture the quasi-periodicity of the index integrand:
\be
\label{eq:Q_ishift}
Q_i(u;\xi,\nu_R,\omega) \, \cZ(u;\xi,\nu_R,a\omega,b\omega)=\cZ(u-\delta_iab\omega;\xi,\nu_R,a\omega,b\omega) \;,
\ee
valid $\forall\,i$ and where $\delta_i = (\delta_{ij})_{j=1}^{\text{rk}(G)}$ so that $(u-\delta_iab\omega)_j=u_j-\delta_{ij}ab\omega$. 

Because of the double-periodicity of $Q_i$, the actual number of solutions $\hat u_i$ to the system of BAEs \eqref{eq:BAEs} is infinite. However, the solutions can be grouped into a finite number of equivalence classes $[\hat{u}_i]$ such that $\hat{u}_i\sim\hat{u}_i+1\sim\hat u_i+\omega$. In other words, the equations and their solutions are well-defined on a torus $\bT^{2\rk(G)}$ which is the product of $\rk(G)$ identical complex tori of modular parameter $\omega$, and the number of solutions on the torus is finite. The modular invariance \eqref{eq:Q_i modularity} confirms that the equations are well-defined on the torus. We define
\be
\label{eq:MBAE}
\fM_\text{BAE} = \biggl\{[\hat{u}_i]\,,\,\,i=1,\dots,\rk(G) \,\,\Big|\,\, Q_i \bigl( [\hat{u}];\xi,\nu_R,\omega \bigr) = 1\,,\quad w \cdot [\hat{u}] \neq [\hat{u}] \quad \forall\, w \in \cW_G \biggr\}
\ee
as the set of solutions (on the torus) that are not fixed by non-trivial elements of the Weyl group. For definiteness we can choose, as representatives, the elements living in a fundamental domain of the torus with modulus $\omega$, \ie~with $0 \leq \RE \hat u_i<1$ and $0 \leq \IM \hat u_i<\IM \omega$. Notice that, because of (\ref{eq:Q_i modularity}), the solutions must organize into representations of $SL(2,\bZ)$.

As we prove below, thanks to the properties of the BA operators, we can rewrite the superconformal index as a sum over solutions to the BAEs in the following way:
\be
\label{eq:BAEformula}
\cI(p,q;v)=\kappa_G \sum_{\hat{u} \,\in\, \fM_\text{BAE}}\cZ_\text{tot}(\hat{u};\xi,\nu_R,a\omega,b\omega) \; H(\hat{u};\xi,\nu_R,\omega)^{-1} \;.
\ee
Here
\be
\label{eq:Ztot}
\cZ_\text{tot}(u;\xi,\nu_R,a\omega,b\omega) = \sum_{\{m_i\}=1}^{ab}\cZ(u-m\omega;\xi,\nu_R,a\omega,b\omega) \;,
\ee
where $\cZ$ is precisely the integrand defined in \eqref{eq:Z} and
\be
\label{eq:H}
H(u;\xi,\nu_R,\omega) = \underset{ij}{\det}\Biggl[\frac{1}{2\pi i} \, \frac{\partial Q_i(u;\xi,\nu_R,\omega)}{\partial u_j}\Biggr]
\ee
is the contribution from the Jacobian of the change of variables $u_i\mapsto Q_i(u)$. Notice that both the function $H$, and the function $\cZ_\text{tot}$ evaluated on the solutions to the BAEs, are doubly-periodic on the product of complex tori of modular parameter $\omega$.

A specialization of this formula to the case $\tau = \sigma$ was derived in \cite{Closset:2017bse}, while a three-dimensional analog was derived in \cite{Closset:2018ghr}. In the next Section we will spell out in detail how the BA formula uniquely fixes the index for all values of the complex fugacities, using either holomorphy or continuity. In Section~\ref{sec: proof} we will derive the final formula \eqref{eq:BAEformula}, starting from the integral representation \eqref{eq:SCIu}. The proof is rather technical and it does not give new physical insights on the main result. Therefore, uninterested readers may stop here.

\subsection{Continuation to generic fugacities}
\label{sec: continuation}

Our BA formula \eqref{eq:BAEformula} can only be applied for special values of the angular fugacities that satisfy (\ref{condition on fugacities}). We will offer two arguments, one based on holomorphy and the other based on just continuity, that this is enough to completely determine the index for all values of the complex fugacities.

Using the standard definition (\ref{eq:traceindex}), the index is not a single-valued function of the angular fugacities $p,q$---unless the R-charges of chiral multiplets are all even. This is also apparent from the integral formula (\ref{eq:SCI}). On the other hand, regarded as a function of chemical potentials $\tau,\sigma$ each living on the upper half-plane $\bH$, the index is single-valued and holomorphic. Keeping the flavor fugacities fixed in the argument that follows, the BA formula applies to points $(\tau,\sigma) \in \bH^2$ such that $\tau/\sigma \in \bQ_+$. Such a set is dense in a hyperplane $\cJ \cong \bR^3$ of real codimension one in $\bH^2$ defined as $\cJ = \bigl\{ (\tau,\sigma) \, \big| \, \tau/\sigma \in \bR_+ \bigr\}$. Thus, the BA formula determines the index on $\cJ$ by continuity. On the other hand, we know that the index is a holomorphic function on $\bH^2$, therefore its restriction to $\cJ$ completely fixes the function on $\bH^2$ by analytic continuation.

It turns out that we can refine the argument in such a way that we only use continuity, and not holomorphy, of the index. This is because if we think in terms of angular fugacities $p,q$ each living in the open unit disk $\bD$, then the set of points $(p,q) \in \bD^2$ such that $q^a = p^b$ for coprime $a,b\in \bN$ is dense in $\bD^2$. This fact is not completely obvious, and we show it in Appendix~\ref{sec: dense}.

Unfortunately, the index (\ref{eq:SCI}) is not a single-valued function of $p,q$ if we keep the flavor fugacities $v_\alpha$ fixed, unless the R-charges are all even. However, it is always possible to find a change of variables which expresses $\cI$ as a single-valued function of a set of new fugacities. The latter is defined by
\be
\Delta_a = \omega_a(\xi) + r_a\nu_R \qquad\Rightarrow\qquad y_a = e^{2\pi i \Delta_a}=v^{\omega_a}(pq)^{\frac{r_a}{2}} \;,\qquad\quad \forall\,a=1,\dots,n_\chi \;.
\ee
This gives us a set of (redundant) chemical potentials $\Delta_a$, one for each chiral multiplet present in the theory, which must satisfy some linear constraint, following the requirement of invariance of the theory under flavor and R-symmetry. Suppose, indeed, the theory has a superpotential given by
\be
\label{eq:superpot}
W(\Phi)=\sum\nolimits_A W_A(\Phi) \;,
\ee
where each $W_A(\Phi)$ is a gauge-invariant homogeneous polynomial of degree $n_A$. Then, for each term in \eqref{eq:superpot}, the following 
linear constraints must be satisfied:
\be
\label{eq:Wconstraint}
\sum_{a\in A}r_a=2 \;,\qquad\qquad \sum_{a\in A}\omega_a^\alpha=0 \;,\qquad \forall\,\alpha=1,\dots,\text{rk}(G) \;,
\ee
where we used $a\in A$ to indicate the chiral components $\Phi_a$ which are present in $W_A$. The first equation imposes that the superpotential has R-charge 2. The second equation constrains $W$ to be invariant under $G_F$. Indeed, $\omega_a=(\omega_a^\alpha)_{\alpha=1}^{\text{rk}(G_F)}$ are the flavor 
weights carried by $\Phi_a$. A similar role is played by ABJ anomalies.

Translating \eqref{eq:Wconstraint} to the definition of $\Delta_a$, we obtain
\be
\label{eq:Deltaconstraints}
\sum_{a\in A}\Delta_a = 2\nu_R=\tau+\sigma \qquad\quad \forall\,A \;.
\ee
In such a new set of variables we have
\be
\label{eq:ZDelta}
\cZ(u;\Delta,\tau,\sigma) = 
\frac{\prod_{a=1}^{n_\chi}\prod_{\rho_a\in\fR_a}\widetilde{\Gamma}(\rho_a(u)+\Delta_a;\tau,\sigma)}{\prod_{\alpha\in\Delta}\widetilde{\Gamma}(\alpha(u);\tau,\sigma)} \;,
\ee
showing that the index is now a well-defined, single-valued and continuous function (in fact, also holomorphic) of the fugacities $p,\,q,\,y_a$. Indeed, recall that the elliptic gamma function is a single-valued function of its arguments, and notice that the constraints \eqref{eq:Deltaconstraints} always involve integer combinations of $\tau$, $\sigma$, thus never introducing non-trivial monodromies under integer shifts. Once again, the BA formula can be applied whenever $q^a = p^b$ and for generic values of $y_a$. Since such a set of points is dense in the space of generic fugacities, we conclude that the BA formula fixes the index completely.

\subsection{Proof of the formula}
\label{sec: proof}

We prove the formula \eqref{eq:BAEformula} in three steps. First we verify the properties \eqref{eq:Q_iperiodicity} and \eqref{eq:Q_ishift} of the BA operators. Then we use them to modify the contour of the integral \eqref{eq:SCIu} and to reduce it to a sum of simple residues. Finally we prove that the only poles that contribute to the residue formula are determined by the BAEs, thus obtaining \eqref{eq:BAEformula}.

\subsubsection{Properties of the BA operators}

First, we prove the identities \eqref{eq:Q_iperiodicity} and \eqref{eq:Q_ishift}. For later convenience, let us briefly recall the anomaly cancellation conditions that are required to have a well-defined four-dimensional theory. These requirements can be expressed in terms of the anomaly coefficients. In particular, let $\bfi=(i,\alpha)$ collectively denote the Cartan indices of the gauge $\times$ flavor group, where $i=1,\dots,\text{rk}(G)$ are the gauge indices and $\alpha=1,\dots,\text{rk}(G_F)$ are the flavor indices. Moreover, define $\bfa=(a,\rho_a)$ as running over all chiral multiplets components, where $\rho_a$ are the weights of the gauge representation $\fR_a$. Then the anomaly coefficients for gauge/flavor symmetries are defined by
\be
\label{eq:anomalycoeffs}
\cA^{\bfi\bfj\bfk} = \sum_{\bfa} Q_\bfa^\bfi \, Q_\bfa^\bfj \, Q_\bfa^\bfk \;,\qquad\qquad
\cA^{\bfi\bfj} = \sum_{\bfa} Q_\bfa^\bfi \, Q_\bfa^\bfj \;,\qquad\qquad
\cA^{\bfi} = \sum_{\bfa} Q_\bfa^\bfi \;,
\ee
where
$Q_\bfa^\bfi = Q^\bfi_{(a,\rho_a)} = (\rho_a^i,\omega_a^\alpha)$ are the components of the gauge $\times$ flavor weights carried by the chiral multiplets. The first and the last coefficient in \eqref{eq:anomalycoeffs} are associated with the gauge$^3$ and mixed gauge-gravitational$^2$ perturbative anomalies. The second term---sometimes called pseudo-anomaly coefficient---describes the non-perturbative or global anomaly \cite{Witten:1982fp, Elitzur:1984kr, Zhang:1987pw} when the corresponding perturbative anomaly vanishes.

Similarly, the perturbative anomaly coefficients involving the R-symmetry are defined by
\bea
\label{eq:pertRanomalycoeffs}
\cA^{\bfi \bfj R} &= \sum_{\bfa} Q^\bfi_\bfa \, Q^\bfj_\bfa (r_\bfa-1)+\delta^{\bfi\bfj,ij}\sum_{\alpha\in\Delta}\alpha^i\alpha^j \qquad& \cA^{\bfi R R} &= \sum_{\bfa} Q^\bfi_\bfa (r_\bfa-1)^2 \\[0.3em]
\cA^{R R R} &= \sum_{\bfa} (r_\bfa-1)^3+\dim G  & \cA^R &= \sum_\bfa (r_\bfa-1)+\dim G \;,
\eea
whereas the pseudo R-anomaly coefficients are
\be
\label{eq:nonpertRanomalycoeffs}
\cA^{\bfi R} = \sum_{\bfa}Q^\bfi_\bfa (r_\bfa-1)\qquad\qquad \cA^{R R} = \sum_{\bfa} (r_\bfa-1)^2+\dim G \;.
\ee

Anomaly cancellation is realized by a set of conditions on the coefficients defined above, that a well-defined quantum gauge theory must satisfy. We will also restrict to the case that the gauge group $G$ is semi-simple. The conditions for the cancellation of the gauge and gravitational anomaly are
\be
\label{eq:gaugeanomaly}
\cA^{ijk}=\cA^i=0\qquad\text{ and }\qquad \cA^{ij}\in 4\nZ \qquad \text{for $G$ semi-simple} \;.
\ee
The conditions for the cancellation of the ABJ anomalies of $G_F$ and $U(1)_R$, namely that those are global symmetries of the quantum theory, are
\be
\label{eq:flavorRanomaly}
\cA^{ij\alpha} = \cA^{ijR} = 0 \;.
\ee
Finally,
\be
\label{eq:semisimpleanomaly}
\cA^{i\alpha\beta} = \cA^{i\alpha R} = \cA^{i RR} = 0 \qquad\text{and}\qquad \cA^{i\alpha} = \cA^{iR} = 0
\ee
simply follow from the restriction to semi-simple gauge group $G$.

We now focus on describing some properties of the basic BA operator
\be
\label{eq:defbasicBAEop}
P(u;\omega)=\frac{\ds e^{-\pi i\frac{ u^2}{\omega}+\pi i u}}{\theta_0(u;\omega)} \;.
\ee
First, consider the function
\be
\theta_0(u;\omega)=(z;h)_\infty(z^{-1}h;h)_\infty=\prod_{k=0}^\infty(1-zh^k)(1-z^{-1}h^{k+1}) \;,\qquad z=e^{2\pi i u} \;,\; h=e^{2\pi i \omega}
\ee
which is holomorphic in $z$ and $h$, and satisfies the following properties:
\bea
\label{eq:theta_0periodicities}
&\theta_0(u+n+m\omega;\omega) = (-1)^{m} \, e^{-2\pi i mu - \pi i m(m-1)\omega} \, \theta_0(u;\omega)\qquad\quad \forall\,n, m\in\nZ\\[0.3em]
&\theta_0(-u;\omega) = \theta_0(u+\omega;\omega) = -e^{-2\pi iu} \, \theta_0(u;\omega) \;.
\eea
They immediately imply
\bea
\label{eq:Pperiodicities}
& P(-u;\omega)=-P(u;\omega) \\[0.2em]
& P(u+n+m\omega;\omega) = (-1)^{n+m} \, e^{-\frac{\pi i }{\omega}(2 n u+n^2)} \, P(u;\omega) \qquad\qquad \forall\,n,m\in\nZ \;.
\eea
It turns out that the basic BA operator has also nice modular transformation properties:
\be
\label{eq:Pmodularity}
P(u; \omega + 1) = e^{\pi i \frac{u^2}{\omega(\omega+1)}} \, P(u; \omega) \;,\qquad
P\left( \frac u\omega; - \frac1\omega \right) = e^{\pi i \left( \frac{u^2}\omega - \frac\omega6 - \frac1{6\omega} + \frac12 \right)} \, P(u; \omega) \;.
\ee
In order to prove \eqref{eq:Q_ishift}, we also need to show that
\begin{multline}
\label{eq:PmGamma}
P(u+r\nu_R;\omega)^m \; \widetilde{\Gamma}(u+r\nu_R;a\omega,b\omega) = (-1)^{\frac{abm^2}{2}+\frac{m(a+b-1)}{2}} \, e^{-\frac{\pi imu^2}{\omega}+\pi i abm^2u - \pi i m(a+b)(r-1)u} \times {} \\[0.2em]
{} \times h^{-\frac{m^3ab}{6}+\frac{ab(a+b)m^2(r-1)}{4}-\frac{m(a+b)^2(r-1)^2}{8}+\frac{m(a^2+b^2+2)}{24}} \; \widetilde{\Gamma}(u+r\nu_R-mab\omega;a\omega,b\omega) \;.
\end{multline}
Here $r\in\nR$ mimics the contribution from the R-charge of a generic multiplet in the theory. Notice that all factors in front of $\widetilde{\Gamma}$ in the r.h.s. of \eqref{eq:PmGamma} explicitly depend on the fermion R-charge $r-1$. This will be crucial to ensure anomaly cancellation in the full BA operator.

\begin{proof}
The identity \eqref{eq:PmGamma} follows from the properties of the elliptic gamma function. Indeed, for generic $\tau$ and $\sigma$, we have that
\be
\label{eq:Gammaperiodicities}
\widetilde{\Gamma}(u+\tau;\tau,\sigma) = \theta_0(u;\sigma) \,\widetilde{\Gamma}(u;\tau,\sigma) \;,\qquad\qquad \widetilde{\Gamma}(u+\sigma;\tau,\sigma) = \theta_0(u;\tau) \, \widetilde{\Gamma}(u;\tau,\sigma) \;.
\ee
Moreover, there exists a factorization property (see Theorem 5.4 of \cite{math/9907061}) which expresses $\widetilde{\Gamma}(u;a\omega,b\omega)$ as a product of elliptic gamma functions with equal periods:
\be
\label{eq:Gammafactorization}
\widetilde{\Gamma}(u;a\omega,b\omega)=\prod_{r=0}^{a-1}\prod_{s=0}^{b-1}\widetilde{\Gamma}(u+(as+br)\omega;ab\omega,ab\omega) \;,
\ee
valid for $a,b\in\nZ$ (not necessarily coprime). Using both \eqref{eq:Gammaperiodicities} and \eqref{eq:Gammafactorization} we obtain the identity
\be
\label{eq:Gammaperiodicity}
\widetilde{\Gamma}(u+ab\omega;a\omega,b\omega)= \left[ \; \prod_{r=0}^{a-1}\prod_{s=0}^{b-1} \theta_0 \big( u+(as+br)\omega;ab\omega \big) \, \right] \times\widetilde{\Gamma}(u;a\omega,b\omega)
\ee
and its generalizations to $m\in\nZ$, given by
\begin{multline}
\label{eq:Gammaperiodicitym}
\widetilde{\Gamma}(u+mab\omega;a\omega,b\omega) = (-z)^{-\frac{abm(m-1)}{2}} \; h^{-\frac{m(m-1)}{2}\frac{ab(2ab-a-b)}{2}-\frac{m(m-1)(m-2)a^2b^2}{6}} \times {} \\[0.2em]
{} \times \left[ \; \prod_{r=0}^{a-1}\prod_{s=0}^{b-1}\theta_0 \big( u+(as+br)\omega;ab\omega \big)^m \, \right] \times\widetilde{\Gamma}(u;a\omega,b\omega) \;.
\end{multline}
Now, by enforcing the assumption that $\text{gcd}(a,b)=1$, we can use the properties of numerical semigroups (see Appendix~\ref{sec:Frobenius} for more details) to reduce the periods of the theta functions from $ab\omega$ to $\omega$. In order to do so, let us introduce some notation. We call $\cR(a,b)$ the set of non-negative integer linear combinations of $a, b$:
\be
\cR(a,b)=\{am+bn\,\vert\,m,n\in\nZ_{\ge0}\}.
\ee
Then $\cR(a,b)$ forms a numerical semigroup, which can be thought of as a subset of $\nZ_{\ge0}$, closed under addition, with only a finite number of excluded non-vanishing elements. The latter elements form the so-called set of gaps $\overline{\cR}(a,b) = \nN\setminus\cR(a,b)$. The highest element of $\overline{\cR}(a,b)$ is the Fr\"obenius number $F(a,b)=ab-a-b$, whereas the order of $\overline{\cR}(a,b)$ is called the genus $\chi(a,b)$ and the sum of all its elements is the weight $w(a,b)$. It is a classic result in mathematics that, in terms of $a,b$, the latter read
\be
\chi(a,b)=\frac{(a-1)(b-1)}{2} \;,\qquad\qquad w(a,b) = \frac{(a-1)(b-1)(2ab-a-b-1)}{12} \;.
\ee
Thanks to the properties of these objects, we can use the following identities (proved in Appendix~\ref{sec:Frobenius}):
\bea
\label{eq:pochidentities}
\prod_{r=0}^{a-1}\prod_{s=0}^{b-1}(zh^{as+br};h^{ab})_\infty &= \frac{(z;h)_\infty}{\prod_{k\in\overline{\cR}(a,b)}(1-zh^k)}\\[0.2em]
\prod_{r=0}^{a-1}\prod_{s=0}^{b-1}(z^{-1}h^{ab-as-br};h^{ab})_\infty &= (z^{-1}h;h)_\infty\prod_{k\in\overline{\cR}(a,b)}(1-z^{-1}h^{-k}) \;,
\eea
which lead to
\be
\label{eq:theta_0reduced}
\prod_{r=0}^{a-1} \prod_{s=0}^{b-1} \theta_0 \big( u+(as+br)\omega;\omega \big) = (-z)^{-\chi(a,b)} \; h^{-w(a,b)} \; \theta_0(u;\omega) \;.
\ee
Substituting into \eqref{eq:Gammaperiodicitym} we obtain
\begin{multline}
\label{eq:Gammathetaperiodicity}
\widetilde{\Gamma}(u+mab\omega;a\omega,b\omega) = (-z)^{-\frac{abm^2}{2}+\frac{m(a+b-1)}{2}} \times {} \\[0.2em]
{} \times h^{-\frac{abm^3}{6}+\frac{ab(a+b)m^2}{4}-\frac{(a^2+b^2+3ab-1)m}{12}} \;
\theta_0(u;\omega)^m \; \widetilde{\Gamma}(u;a\omega,b\omega) \;.
\end{multline}
Finally, applying \eqref{eq:Gammathetaperiodicity} to the l.h.s. of \eqref{eq:PmGamma} proves the latter identity.
\end{proof}

We now turn to analyzing the full BA operators. Notice that, in the definition \eqref{eq:Q_i}, $Q_i$ receive contribution only from the chiral multiplets of the theory. The vector multiplets do not appear in \eqref{eq:Q_i} because their contribution simply amounts to
\be
\label{eq:Palpha}
\prod_{\alpha\in\Delta} P\big( \alpha(u);\omega \big)^{-\alpha^i} = \prod_{\alpha>0} \biggl[\frac{P\big( -\alpha(u);\omega \big)}{P\big(\alpha(u);\omega\big)}\biggr]^{\alpha^i}=(-1)^{\sum_{\alpha>0}\alpha^i}=1 \;,
\ee
which holds true if $G$ is semi-simple, as in this case the sum of positive roots is always an even integer. Despite this fact, as far as the proof of \eqref{eq:BAEformula} is concerned, we find it more convenient to write
\bea
\label{eq:Q_i2}
Q_i(u;\xi,\nu_R,\omega) = \prod_{a=1}^{n_\chi}\prod_{\rho_a\in\fR_a}P\big( \rho_a(u)+\omega_a(\xi)+r_a\nu_R;\omega\big)^{\rho_a^i} \times \prod_{\alpha\in\Delta}P\big( \alpha(u);\omega \big)^{-\alpha^i}
\eea
without simplifying the vector multiplet contribution.

At this point, using \eqref{eq:Pperiodicities} we can show that $Q_i$ satisfy:
\bea
Q_i(u+n;\xi,\nu_R,\omega) &= (-1)^{\cA^{ij}n_j} \, e^{-\frac{\pi i}{\omega} \left( \cA^{ijk}n_j(2u_k+n_k)+2\cA^{ij\alpha}n_j\xi_\alpha+2\cA^{ijR}n_j\omega \right)} \,Q_i(u;\xi,\nu_R,\omega)\\[0.2em]
Q_i(u+m\omega;\xi,\nu_R,\omega) &= (-1)^{\cA^{ij}m_j} \, Q_i(u;\xi,\nu_R,\omega) \;,
\eea
which, thanks to \eqref{eq:gaugeanomaly}--\eqref{eq:semisimpleanomaly}, reduce to \eqref{eq:Q_iperiodicity} $\forall\,n_i,m_i\in\nZ$ in an anomaly-free theory. Similarly, \eqref{eq:Pperiodicities} and \eqref{eq:Pmodularity} together with the anomaly cancelation conditions \eqref{eq:gaugeanomaly}--\eqref{eq:semisimpleanomaly} imply \eqref{eq:Q_i modularity}. Moreover, using \eqref{eq:Q_i2}, we can write
\begin{multline}
Q_i(u;\xi,\nu_R,\omega) \; \cZ(u;\xi,\nu_R,a\omega,b\omega) = {} \\[0.3em]
{} = \frac{\prod_{a,\rho_a} P\big( \rho_a(u)+\omega_a(\xi)+r_a\nu_R;\omega \big)^{\rho_a^i} \; \widetilde{\Gamma}\big( \rho_a(u)+\Delta_a;a\omega,b\omega \big)}{\prod_{\alpha\in\Delta} P\big( \alpha(u);\omega\big)^{\alpha^i} \; \widetilde{\Gamma}\big( \alpha(u);a\omega,b\omega \big)} \;.
\end{multline}
Applying \eqref{eq:PmGamma}, the latter equation reduces to
\begin{multline}
Q_i(u;\xi,\nu_R,\omega) \; \cZ(u;\xi,\nu_R,a\omega,b\omega) = (-1)^{\frac{ab}{2}\cA^{ii}+\frac{a+b-1}{2}\cA^i} \; e^{\pi i ab \left( \cA^{iij}u_j+\cA^{ii\alpha}\xi_\alpha \right)}\times {} \\[0.3em]
{} \times e^{-\frac{\pi i}{\omega} \left( \cA^{ijk}u_ju_k+\cA^{i\alpha\beta}\xi_\alpha\xi_\beta+2\cA^{ij\alpha}u_j\xi_\alpha \right)} \; e^{-\pi i(a+b) \left( \cA^{ijR}u_j+\cA^{i\alpha R}\xi_\alpha \right) + \frac{\pi iab(a+b)}{2}\cA^{iiR}\omega} \times {} \\[0.3em]
{} \times e^{-\frac{\pi i(a+b)^2}{4}\cA^{iRR}\omega-\frac{\pi ia^2b^2}{3}\cA^{iii}\omega+\frac{\pi i(a^2+b^2)}{12}\cA^i\omega} \; \cZ(u-\delta_iab\omega;\xi,\nu_R,\omega) \;,
\end{multline}
which, by anomaly cancellation, reduces to \eqref{eq:Q_ishift}.

\subsubsection{Residue formula}

We now use the BA operators and their properties to modify the contour of integration of the index in \eqref{eq:SCIu}. For our purposes, it is sufficient to implement the following trivial relation:
\bea
\label{eq:Qintheindex}
\cI(p,q;v)&=\kappa_G \oint\cZ(u;\xi,\nu_R,a\omega,b\omega) \; d^{\text{rk}(G)}u \\[0.2em]
	&= \kappa_G \oint \frac{\prod_{i=1}^{\text{rk}(G)}\big(1-Q_i(u;\xi,\nu_R,\omega)\big)}{\prod_{i=1}^{\text{rk}(G)}\big(1-Q_i(u;\xi,\nu_R,\omega)\big)} \; \cZ(u;\xi,\nu_R,a\omega,b\omega) \; d^{\text{rk}(G)}u \;.
\eea
The numerator of the integrand can be expanded as
\bea
\label{eq:QZ}
\prod_{i=1}^{\text{rk}(G)}\bigl(&1-Q_i(u;\xi,\nu_R,\omega)\bigr) \times \cZ(u;\xi,\nu_R,a\omega,b\omega) = {} \\[0.2em]
&=\sum_{n=0}^{\text{rk}(G)}\frac{(-1)^n}{n!} \sum_{i_1\neq \dots\neq i_n}^{\text{rk}(G)}Q_{i_1}(u;\xi,\nu_R,\omega) \ldots Q_{i_n}(u;\xi,\nu_R,\omega)\,\cZ(u;\xi,\nu_R,a\omega,b\omega)\\[0.2em]
&=\sum_{n=0}^{\text{rk}(G)}\frac{(-1)^n}{n!} \sum_{i_1\neq\dots\neq i_n}^{\text{rk}(G)} \cZ\bigl(
u - (\delta_{i_1} + \ldots + \delta_{i_n})ab\omega;\xi,\nu_R,a\omega,b\omega\bigr) \;,
\eea
where, in the last line, we have used the shift property \eqref{eq:Q_ishift}. Plugging the last equation back in \eqref{eq:Qintheindex} gives:
\be
\label{eq:Iwithi_1i_n}
\cI(p,q;v)=\kappa_G\sum_{n=0}^{\text{rk}(G)}\frac{(-1)^n}{n!}\sum_{i_1\neq\dots\neq i_n}^{\text{rk}(G)} I_{i_1\dots i_n}(p,q;v) \;,
\ee
with
\bea
\label{eq:Ii_1i_n}
I_{i_1\dots i_n}(p,q;v) &= \oint_{\nT^{\rk(G)}}\frac{\cZ\big(u-(\delta_{i_1}+\ldots+\delta_{i_n})ab\omega;\xi,\nu_R,a\omega,b\omega\big)}{\prod_{i=1}^{\rk(G)}\bigl(1-Q_i(u;\xi,\nu_R,\omega)\bigr)} \; d^{\text{rk}(G)}u \\[0.3em]
&= \oint_{\cC_{i_1 \dots i_n}}\frac{\cZ\big(u;\xi,\nu_R,a\omega,b\omega\big)}{\prod_{i=1}^{\rk(G)}\bigl(1-Q_i(u;\xi,\nu_R,\omega)\bigr)} \; d^{\rk(G)}u
\eea
and where
\be
\label{eq:Ci1in}
\cC_{i_1 \dots i_n} = \bT^{\rk(G)-n}\times\bigcup_{k=1}^n\bigl\{\abs{z_{i_k}}=\abs{h}^{-ab}; \circlearrowleft\bigr\} \;.
\ee
This is a contour where $z_{i_1},\dots,z_{i_n}$ live on circles of radius $\abs{h}^{-ab}$, whereas the other variables $z_j$ parametrize the unit circles in $\bT^{\rk(G)-n}$. The second line in \eqref{eq:Ii_1i_n} has been obtained by implementing the change of variables $u_{i_k}\mapsto u_{i_k}+ab\omega$ for $k=1,\dots,n$ and using the periodicity \eqref{eq:Q_iperiodicity}.

The series of integrals in \eqref{eq:Iwithi_1i_n} can be resummed to a unique integral over a composite contour:
\be
\label{eq:IwithC}
\cI(p,q;v)=\kappa_G\oint_{\cC} \frac{\cZ\big(u;\xi,\nu_R,a\omega,b\omega\big)}{\prod_{i=1}^{\rk(G)}\big(1-Q_i(u;\xi,\nu_R,\omega)\big)} \; d^{\rk(G)}u \;,
\ee
where
\be
\label{eq:C}
\cC=\sum_{n=0}^{\rk(G)}\frac{(-1)^n}{n!}\bigcup_{i_1\neq\dots\neq i_n}^{\rk(G)}\cC_{i_1\dots i_n}\simeq\bigcup_{i=1}^{\rk(G)}\bigl\{\abs{z_i}=1; \circlearrowleft\bigr\}\cup\bigl\{\abs{z_i}=\abs{h}^{-ab}; \circlearrowright\bigr\}
\ee
is a contour encircling the annulus $\cA = \left\{ u_i \,\middle\vert\,1<\abs{z_i}<\abs{h}^{-ab},\,i=1,\dots,\rk(G) \right\}$.

We now apply the residue theorem to \eqref{eq:IwithC}. The integrand has simple poles coming from the denominator, whose positions are precisely described by the BAEs \eqref{eq:BAEs}. Obviously, only the poles that lie inside the annulus $\cA$ contribute to the contour integral. Moreover, as we do in Appendix~\ref{app: Weyl}, one can show that whenever a particular solution $[\hat u]$ to the BAEs \eqref{eq:BAEs} is fixed (on the torus) by a non-trivial element of the Weyl group $\cW_G$, namely $w \cdot [\hat u] = [\hat u]$, then the numerator $\cZ(\hat u; \xi, \nu_R, a\omega, b\omega)$ is such that cancelations take place and there is no contribution to the integral---more precisely, the function $\cZ_\text{tot}(\hat u; \xi, \nu_R, a\omega, b\omega)$ defined in (\ref{eq:Ztot}) vanishes.%
\footnote{In particular, let us stress that the condition $w\cdot[\hat u] \neq [\hat u]$ in the definition of $\fM_\text{BAE}$ could be relaxed with no harm: in that case, we would simply include more poles in the sum, whose residues however combine to zero.}
Hence, we define the set of relevant poles by:
\be
\label{eq:Mindexold}
\fM_\text{index} = \Bigl\{\hat{u}_i \;\Big\vert\; [\hat u_i] \in \fM_\text{BAE} \quad\text{and}\quad 1<\abs{\hat{z_i}}<\abs{h}^{-ab}\,,\; i=1,\dots,\rk(G)\Bigr\} \;.
\ee
This includes all points inside the annulus $\cA$ such that their class belongs to $\fM_\text{BAE}$. In particular, the same equivalence class $[\hat{u}_i]\in\fM_\text{BAE}$  appears in $\fM_\text{index}$ as many times as the number of its representatives living in $\cA$. For this reason, we employ the following alternative description:
\be
\label{eq:Mindex}
\fM_\text{index}= \Bigl\{\hat{u}^{\scriptscriptstyle(\vec{m})}_i=[\hat{u}_i]-m_i\,\omega \,\,\Big\vert\,\, [\hat{u}_i]\in\fM_\text{BAE}\,, \; m_i=1,\dots,ab \,, \; i=1,\dots,\rk(G)\Bigr\}
\ee
where, we some abuse of notation, we have denoted as $[\hat u_i]$ the representative in the fundamental domain of the torus as after \eqref{eq:MBAE}.

In addition, the numerator $\cZ$ has other poles coming from the elliptic gamma functions. As we show below, as long as the fugacities $v_\alpha, p, q$ are taken within the domain \eqref{eq:domaininv}---which is necessary in order for the standard contour integral representation \eqref{eq:SCI} to be valid---those other poles either lie outside the annulus $\cA$ or are not poles of the integrand (because the denominator has a pole of equal or higher degree) and thus do not contribute to the integral.

Therefore, working within the domain \eqref{eq:domaininv}, we can rewrite the index as
\be
\label{eq:Iresidues}
\cI(p,q;v) = (-2\pi i)^{\rk(G)} \; \kappa_G \sum_{\hum \in \, \fM_\text{index}} \Res{u=\hum} \Biggl[ \; \frac{\cZ(u;\xi,\nu_R,a\omega,b\omega)}{\prod_{i=1}^{\rk(G)}\bigl(1-Q_i(u;\xi,\nu_R,\omega)\bigr)} \, d^{\rk(G)}u \, \Biggr] \;.
\ee
Computing the residues produces the final expression for the supersymmetric index:
\be
\label{eq:IBAEs}
\cI(p,q;v) = \kappa_G \sum_{\hum \in \, \fM_\text{index}} \cZ(\hum;\xi,\nu_R,a\omega,b\omega) \; H(\hum;\xi,\nu_R,\omega)^{-1} \;,
\ee
where $H$ is defined in \eqref{eq:H}. The residue formula \eqref{eq:IBAEs} can be rewritten, more elegantly, in the final form:
\be
\label{eq:finalformula}
\cI(p,q;v)=\kappa_G\sum_{\hat{u} \,\in\, \fM_\text{BAE}} \cZ_\text{tot}(\hat{u};\xi,\nu_R,a\omega,b\omega) \; H(\hat{u};\xi,\nu_R,\omega)^{-1} \;,
\ee
where
\be
\label{eq:Ztot2}
\cZ_\text{tot}(u;\xi,\nu_R,a\omega,b\omega) = \sum_{\{m_i\}=1}^{ab}\cZ(u-m\omega;\xi,\nu_R,a\omega,b\omega) \;.
\ee
To obtain this expression we have split the sum over the poles in $\fM_\text{index}$ into a sum over the inequivalent solutions to the BAEs, described by the elements of $\fM_\text{BAE}$, and the sum over the ``repetitions'' of these elements in the annulus $\cA$. Moreover, we have used the double-periodicity of the Jacobian $H(u;\xi,\nu_R,\omega)$ to pull the latter sum inside the definition of $\cZ_\text{tot}$.

\subsubsection{Analysis of the residues}

The last step consists in showing that the only residues contributing to \eqref{eq:IwithC} come from zeros of the denominator. In particular we need to show that, remaining within the domain \eqref{eq:domaininv}, all poles in \eqref{eq:IwithC} which are not given by the BAEs live outside the annulus $\cA$ and thus do not contribute to the integral. We concretely do so by proving that every pole of $\cZ$ inside $\cA$ is also a pole of the denominator $\prod_i(1-Q_i)$ with a high enough degree that the integrand of \eqref{eq:IwithC} is non-singular at those points. 

We begin by classifying the poles of $\cZ$. Using \eqref{eq:theta_0periodicities}, \eqref{eq:Gammaperiodicities} and
\be
\widetilde{\Gamma}(u;\tau,\sigma)=\frac{1}{\widetilde{\Gamma}(\tau+\sigma-u;\tau,\sigma)} \;,
\ee
we can rewrite $\cZ$ as
\be
\label{eq:thetaZ}
\cZ(u;\xi,\nu_R,a\omega,b\omega) = \prod_{\alpha>0} \theta_0 \bigl( \alpha(u);a\omega \bigr) \, \theta_0 \bigl( -\alpha(u);b\omega \bigr) \times \prod_{a, \rho_a}\widetilde{\Gamma} \bigl( \rho_a(u)+\omega_a(\xi)+r_a\nu_R;a\omega,b\omega \bigr) \;.
\ee
Since $\theta_0(u;\omega)$ has no poles for finite $u$, the only singularities of $\cZ$ come from the elliptic gamma functions related to the chiral multiplets. These can be read off the product expansion:
\be
\widetilde{\Gamma}(u;a\omega,b\omega)=\prod_{m=0}^\infty\prod_{r=0}^{a-1}\prod_{s=0}^{b-1}\biggl(\frac{1-h^{ab(m+2)-as-br}z^{-1}}{1-h^{abm+as+br}z}\biggr)^{m+1}
\ee
that follows from \eqref{eq:Gammafactorization}, and so they are given by
\be
\label{eq:polesZ}
z^{\rho_a}=v^{-\omega_a}h^{-r_a(a+b)/2-abm-as-br}
\ee
for $0\le r\le a-1$, $0\le s\le b-1$ and $m\ge0$. The multiplicity of each pole is $\mu^a_m=m+1$.%
\footnote{In counting the multiplicity one may worry that there could be different choices of $r, s$ that give the same $abm+as+br$ for fixed $m$. This is equivalent to finding non-trivial solutions to the equation $as+br=as'+br'$. However, it is easy to see that, as long as $0\le r, r'\le a-1$ and $0\le s, s'\le b-1$, such an equation has no non-trivial solution in $\nZ$.}
Notice that one could also write $z^{\rho_a}=v^{-\omega_a}h^{-r_a(a+b)/2-k}$ for $k\in\cR(a,b)$.

We now turn to analyzing the denominator. More specifically, we need to find the singularities of $\prod_{i=1}^{\rk(G)} \bigl( 1-Q_i(u) \bigr)$. From \eqref{eq:Q_i} and \eqref{eq:basicBAEop} we see that $Q_i$ has a pole whenever $\theta_0 \bigl( \rho_a(u)+\omega_a(\xi)+r_a\nu_R;\omega \bigr)=0$ and $\rho_a^i>0$. Therefore, the singularities of the denominator are given by
\be
\label{eq:zerosQ_i}
z^{\rho_a}=v^{-\omega_a}h^{-r_a(a+b)/2+n} \qquad\quad \text{ for }n\in\nZ \;,
\ee
all with the same multiplicity $\nu^a=\sum_{i \,\in\, \fD_a^+}\rho_a^i$. Here $\cD_a^\pm$ represents the set of indices such that $\rho_a^i>0$, resp.~$\rho_a^i<0$, thus $\nu^a$ is the sum of the positive components of $\rho_a$. We notice that the denominator poles in \eqref{eq:zerosQ_i} with $-n\in\cR(a,b)$ coincide with the numerator poles. Therefore, the actual singularities of the integrand in \eqref{eq:IwithC} are only those points in \eqref{eq:polesZ} such that $\mu_m^a>\nu^a$, or more explicitly
\be
\label{eq:multiplicity}
m \ge \sum\nolimits_{i \,\in\, \fD_a^+}\rho_a^i \;.
\ee

We now want to show that, when the fugacities satisfy \eqref{eq:domaininv}, the set of actual singularities is always living outside the annulus $\cA$. Therefore, we first study the conditions for which \eqref{eq:polesZ} belong to the annulus $\cA$. By imposing that $1<\abs{z_i}<\abs{h}^{-ab}$, we obtain that
\be
\label{eq:conditionpolesA}
\abs{h}^{-ab\sum_{i\in\fD_a^-}\rho_a^i}<\abs{z^{\rho_a}}<\abs{h}^{-ab\sum_{i\in\fD_a^+}\rho_a^i} \;, \qquad\quad \forall\,a \;.
\ee
Then we determine the constraints imposed on \eqref{eq:polesZ} by requiring \eqref{eq:domaininv}. In the rational case, the latter conditions are expressed by $ \abs{h}^{a+b}<\abs{v^{\omega_a}h^{r_a(a+b)/2}}<1$, $\forall\,a$. These inequalities, together with $0\le as+br\le 2ab-a-b$, imply that
\be
\abs{h}^{-abm}\le\abs{h}^{-abm-as-br}<\abs{z^{\rho_a}}<\abs{h}^{-abm-a(s+1)-b(r+1)}\le \abs{h}^{-ab(m+2)} \;.
\ee
Furthermore, requiring \eqref{eq:multiplicity} to be satisfied, we obtain that
\be
\label{eq:conditionpoles1}
\abs{z^{\rho_a}}>\abs{h}^{-ab\sum_{i\in\fD_a^+}\rho_a^i} \;, \qquad\quad\forall\,a \;,
\ee
which is satisfied by all the singularities of \eqref{eq:IwithC} coming from the numerator $\cZ$.

At this point, we immediately notice that the intersection between \eqref{eq:conditionpolesA} and \eqref{eq:conditionpoles1} is empty. This means that, if the flavor fugacities satisfy \eqref{eq:domaininv}, all poles of the integrand \eqref{eq:IwithC} that come from poles of the numerator $\cZ$ live outside the annulus $\cA$, and so the only residues contributing to the integral are those given by the BAEs. This completes the proof of \eqref{eq:BAEformula}.



\section*{Acknowledgments} 

F.B. is supported in part by the MIUR-SIR grant RBSI1471GJ ``Quantum Field Theories at Strong Coupling: Exact Computations and Applications''.

\appendix

\section{Numerical semigroups and the Fr\"obenius problem}
\label{sec:Frobenius}

Given a set of non-negative integer numbers $\{a_1,\dots,a_r\}$, the Fr\"obenius problem consists in classifying which integers can (or cannot) be written as non-negative integer linear combinations of those. This problem has deep roots in the theory of numerical semigroups.

A {\itshape semigroup} is an algebraic structure $\cR$ endowed with an associative binary operation. Analogously to groups, we denote it as $(\cR,*)$. On the other hand, differently from the case of a group, no requirement on the presence of identity and inverse elements is made.  A {\itshape numerical semigroup} is an additive semigroup $(\cR,+)$, where $\cR$ consists of all non-negative integers $\nZ_{\ge0}$ except for a finite number of positive elements (thus $0\in\cR$). The set $\{n_1,\dots,n_t\}$ is called a {\itshape generating set} for $(\cR,+)$ if all elements of $\cR$ can be written as non-negative integers linear combinations of $n_1,\dots,n_t$. We then denote the semigroup with the presentation
\be
\cR=\braket{n_1,\dots,n_t}.
\ee
Among all possible presentations of $\cR$, there exists a {\itshape unique minimal presentation}, which contains the minimal number of generators. Such a number is called the {\itshape embedding dimension} $e(\cR)$ of the semigroup. We now define other important quantities associated with numerical semigroups:
\begin{itemize}
	\item The {\itshape multiplicity} $m(\cR)$ is the smallest non-zero element of $\cR$.
	\item The {\itshape set of gaps} $\overline{\cR} = \nN\setminus\cR$ is the set of positive integers which are not contained in $\cR$. Equivalently, the gaps are defined as all natural numbers which cannot be written as non-negative integer linear combination of the generators $n_1,\dots,n_t$ of $\cR$.
	\item The set of gaps $\overline{\cR}$ is always a finite set. Its largest element is the {\itshape Fr\"obenius number} $F(\cR)$. Alternatively, given a presentation $\braket{n_1,\dots,n_t}$, the Fr\"obenius number is defined as the largest integer which cannot be written as a non-negative integer linear combination of the generators.
	\item The {\itshape genus} $\chi(\cR)$ is the number of gaps, \ie~it is the order of the set of gaps: $\chi(\cR)=\Abs{\overline{\cR}}$.
	\item The {\itshape weight} $w(\cR)$ is the sum of all gaps: $w(\cR)=\sum_{k\in\overline{\cR}}k$.
	\item The following inequalities hold:
	\be
	e(\cR)\le m(\cR)\qquad\qquad F(\cR)\le2\chi(\cR)-1.
	\ee
	In particular, if $x\in\cR$, then $F(\cR)-x\notin\cR$.
\end{itemize} 
We now study the case where the embedding dimension is $e(\cR)=2$, \ie~the minimal presentation is defined by two positive integers $a,b$ with $\text{gcd}(a,b)=1$. The associated numerical semigroup is denoted by $\cR(a,b)=\braket{a,b}$ and the set of gaps is $\overline{\cR}(a,b)=\nN\setminus\cR(a,b)$. The multiplicity is simply $m(a,b)=\min\{a,b\}$, whereas the Fr\"obenius number is given by
\be
F(a,b)=ab-a-b \;.
\ee
The genus and the weight are
\be
\chi(a,b)=\frac{(a-1)(b-1)}{2}\qquad\qquad w(a,b)=\frac{(a-1)(b-1)(2ab-a-b-1)}{12} \;.
\ee

Thanks to the properties of $\cR(a,b)$, one can prove the following identities:
\bea
& \prod_{r=0}^{a-1}\prod_{s=0}^{b-1}(zh^{as+br};h^{ab})_\infty = \frac{(z;h)_\infty}{\prod_{k\in\overline{\cR}(a,b)}(1-zh^k)}\\[0.2em]
& \prod_{r=0}^{a-1}\prod_{s=0}^{b-1}(z^{-1}h^{ab-as-br};h^{ab})_\infty = (z^{-1}h;h)_\infty\prod_{k\in\overline{\cR}(a,b)}(1-z^{-1}h^{-k}) \;.
\eea
\begin{proof}
We begin with the first identity. Using the definition of the $q$-Pochhammer symbol we can write:
\be
\prod_{r=0}^{a-1}\prod_{s=0}^{b-1}(zh^{as+br};h^{ab})_\infty=\prod_{n=0}^\infty\prod_{r=0}^{a-1}\prod_{s=0}^{b-1}(1-zh^{abn+as+br}) \;.
\ee
Using that $a,b$ are coprime, the set of integers $\bigl\{ as + br \,\big|\, r = 0, \dots, a-1,\; s = 0, \dots, b-1 \bigr\}$ covers once and only once every class modulo $ab$. It follows that the set of exponents $\{abn + as + br\}$ is precisely $\cR(a,b)$. Then
\be
\prod_{r=0}^{a-1}\prod_{s=0}^{b-1}(zh^{as+br};h^{ab})_\infty = \!\! \prod_{k\in\cR(a,b)} \! (1-zh^{k}) = \frac{\prod_{k=0}^\infty(1-zh^{k})}{\prod_{k\in\overline{\cR}(a,b)}(1-zh^{k})} =\frac{(z;h)_\infty}{\prod_{k\in\overline{\cR}(a,b)}(1-zh^k)} \;,
\ee
which proves the first equality in \eqref{eq:pochidentities}.

The proof of the second identity is a bit trickier. The key point is to notice that the set $\{as + br\}$ does not contain any element of $\wb \cR(a,b)$ and thus
\be
\{as + br\} = \bigl\{ k + \Delta_k ab \,\big|\, k = 0, \dots, ab-1 \bigr\} \qquad\text{with}\qquad \Delta_k = \begin{cases} 0 &\text{if } k \in \cR(a,b) \\ 1 &\text{if } k \in \wb \cR(a,b) \;. \end{cases}
\ee
This implies that $\{ ab - as - br \} = \bigl\{ -k + (1-\Delta_k) ab \,\big|\, k = 0, \dots, ab-1 \bigr\}$. Finally, including the freedom of choosing $n \geq 0$, we find that the set of exponents is
\be
\{ abn + ab - as - ar \} = (-\wb \cR) \cup \bZ_{>0} \;.
\ee
Then
\begin{multline}
\prod_{r=0}^{a-1}\prod_{s=0}^{b-1}(z^{-1}h^{ab-as-br};h^{ab})_\infty = \prod_{n=0}^\infty\prod_{r=0}^{a-1}\prod_{s=0}^{b-1} \bigl( 1-z^{-1}h^{ab(n+1)-as-br} \bigr) \\[0.2em]
{} = \prod_{k \in \wb\cR(a,b)} ( 1 - z^{-1} h^{-k}) \times \prod_{k=1}^\infty (1- z^{-1} h^{k}) = (z^{-1}h;h)_\infty\prod_{k\in\overline{\cR}(a,b)}(1-z^{-1}h^{-k}) \;.
\end{multline}
This completes the proof of \eqref{eq:pochidentities}.
\end{proof}

Thanks to the definition of $\theta_0(u;\omega)$, we can apply \eqref{eq:pochidentities} and we obtain that
\be
\prod_{r=0}^{a-1}\prod_{s=0}^{b-1} \theta_0 \bigl( u+(as+br)\omega;\omega \bigr) = \prod_{k\in\overline{\cR}(a,b)}\frac{(1-z^{-1}h^{-k})}{(1-zh^k)} \; \theta_0(u;\omega)
= \frac1{(-z)^{\chi(a,b)} \; h^{w(a,b)}} \; \theta_0(u;\omega) \;.
\ee

\section{A dense set}
\label{sec: dense}

Here we show that the set of points $(p,q)$ such that
\be
\label{condition on p,q appendix}
q^a = p^b \qquad\text{for coprime $a,b\in\bN$}
\ee
is dense in $\bigl\{ |p|<1,\, |q|<1\bigr\}$. We write the fugacities in terms of chemical potentials, $p = e^{2\pi i \sigma}$ and $q = e^{2\pi i \tau}$ with $\IM \sigma, \IM \tau>0$, and for the sake of this argument we choose the determination on the ``strip'' $0 \leq \RE \sigma, \RE \tau<1$. Then the condition (\ref{condition on p,q appendix}) is equivalent to
\be
\label{condition on tau,sigma appendix}
a(\tau + n) = b(\sigma + m)
\ee
for some $m,n\in \bZ$ and $a,b\in \bN$ coprime.

We choose an arbitrary point $(\tau_0, \sigma_0)$ in the strip and ask if we can find another point $(\tau,\sigma)$, arbitrarily close, that satisfies (\ref{condition on tau,sigma appendix}). Consider a straight line in the complex plane that starts from 0 and goes through $\tau_0 + n$ for some integer $n$. When winding once around the strip, this line has an imaginary excursion
\be
\Delta y = \frac{\IM \tau_0}{\RE \tau_0 + n} \;.
\ee
We can make this quantity arbitrarily small by choosing $n$ sufficiently large.
We define $\sigma'$ as the closest point to $\sigma_0$ that lies on the image of the line on the strip modulo 1, and has $\RE \sigma'=\RE \sigma_0$. It is clear that
\be
|\sigma' - \sigma_0| = \bigl| \IM \sigma' - \IM \sigma_0 \bigr| \leq \Delta y/2 \;,
\ee
and, by construction, $(\sigma ' + m) = t (\tau_0 + n)$ for some $m,n\in \bZ$ and $t\in \bR_+$. We see that $|\sigma'-\sigma_0|$ can be made arbitrarily small by increasing $n$. Next, we approximate $t$ by a fraction $a/b \in \bQ_+$. This, for $a/b$ sufficiently close to $t$, defines a point $\sigma$ in the strip by
\be
(\sigma + m) = \frac ab \, (\tau_0 + n) \;.
\ee
It is clear that $\sigma$ can be made arbitrarily close to $\sigma'$ by approximating $t$ sufficiently well with $a/b$. We have thus found a pair $(\sigma, \tau=\tau_0)$, arbitrarily close to $(\sigma_0,\tau_0)$, that satisfies the constraint (\ref{condition on tau,sigma appendix}).


\section{Weyl group fixed points}
\label{app: Weyl}

In this appendix we prove that $\cZ_\text{tot}(u; \xi, \nu_R, a\omega, b\omega)$ vanishes when evaluated at a point $\hat u$ which is fixed, on a torus of modular parameter $\omega$, by a non-trivial element $w$ of the Weyl group $\cW_G$:
\be
w \cdot [\hat u] = [\hat u] \;.
\ee
This implies that the solutions to the BAEs (\ref{eq:MBAE}) which are fixed points on the torus of
an element of the Weyl group, can be excluded from the set $\fM_\text{BAE}$---as is done in (\ref{eq:MBAE})---because they do not contribute to the BA formula (\ref{eq:BAEformula}) for the superconformal index.

\subsection{The rank-one case}

Let us first consider the case that the gauge group $G$ has rank one, \ie, that $\fg = \fsu(2)$. Then there are only two roots, $\alpha$ and $-\alpha$, and the Weyl group is $\cW_G = \{1, s_\alpha\} \cong \bZ_2$ where $s_\alpha$ is the unique non-trivial Weyl reflection along the root $\alpha$:
\be
s_\alpha(u) = -u \qquad\qquad\forall\; u \in \fh \;.
\ee
We choose a basis element $\{H\}$ for the Cartan subalgebra $\fh$ such that $\rho(H) \equiv \rho \in \bZ$ for any weight $\rho \in \Lambda_\text{weight}$.
In this canonical basis $\alpha=2$ (while the fundamental weight is $\lambda = 1$). The solutions to $s_\alpha \cdot [\hat u] = [\hat u]$ are given by%
\footnote{The integers $p,q$ appearing in this appendix should not be confused with the complex angular fugacities appearing in the rest of the paper.}
\be
\hat u = \frac{p+q\omega}2 \qquad\qquad\text{with}\quad p,q \in \bZ \;.
\ee
Choosing a representative for $[\hat u]$ in the fundamental domain of the torus, the inequivalent solutions are with $p=0,1$ and $q=0,1$.

The representations of $\fs\fu(2)$ are labelled by a half-integer spin $j\in\nN/2$ and their weights are $\rho \in \big\{ \ell \alpha \,\big|\, \ell = -j, -j +1 , \dots, j - 1, j\big\}$. Therefore, exploiting the expression in (\ref{eq:thetaZ}), the function $\cZ$ reduces to
\begin{multline}
\label{eq:ZZrankone}
\cZ(u;\xi,\nu_R,a\omega,b\omega) = {} \\
{} = \theta_0\big(\alpha(u);a\omega\big) \, \theta_0\big( {-\alpha(u)};b\omega\big) \prod_{a} \prod_{\ell_a=-j_a}^{j_a} \widetilde{\Gamma} \big(\ell_a\alpha(u)+\omega_a(\xi)+r_a\nu_R;a\omega,b\omega \big) \;.
\end{multline}
Moreover, the function $\cZ_\text{tot}$ defined in (\ref{eq:Ztot}) is a single sum over $m=1, \dots, ab$.

We want to prove that $\cZ_\text{tot}(\hat u ;\xi,\nu_R,a\omega,b\omega) = 0$. To do that, we construct an involutive map $\gamma: m \mapsto m'$ acting on the set of integers $\{1, \dots, ab\}$ according to
\be
m' = m \mod b \;,\qquad\qquad m' = q-m \mod a \;,
\ee
which define $m'$ uniquely. It will be convenient to introduce the numbers $r,s \in \bZ$ such that $m' = m + sb = q-m + ra$. The map $\gamma$ has the property that
\be
\label{property map gamma}
m' - q/2 = \begin{cases} m - q/2 &\mod b \,, \\ -(m-q/2) & \mod a \,, \end{cases} \quad = m- q/2 + sb = -(m - q/2) + ra \;.
\ee
We will prove that
\be
\label{inversion property of gamma}
\cZ\big( \hat u - m' \omega; \xi, \nu_R, a\omega, b\omega \big) = - \cZ\big( \hat u - m\omega; \xi, \nu_R, a\omega, b\omega\big) \;.
\ee
In particular, the sum over $m$ inside $\cZ_\text{tot}$ splits into a sum over the fixed points of $\gamma$ and a sum over the pairs of values related by $\gamma$. The property (\ref{inversion property of gamma}) guarantees that each term in those sums vanishes, implying that $\cZ_\text{tot}$ vanishes.

Let us adopt the notation
\be
\cZ_m \,\equiv\, \cZ(\hat u-m\omega;\xi,\nu_R,a\omega,b\omega) = \cZ \bigl( p/2 - (m-q/2) \omega;\xi,\nu_R,a\omega,b\omega\bigr) \;.
\ee
We define the vector multiplet and the chiral multiplet contribution, respectively, as
\bea
\cA_m &= \theta_0\bigl(\alpha(p/2)-\alpha(m-q/2)\omega;a\omega\bigr) \, \theta_0\bigl(-\alpha(p/2)+\alpha(m-q/2)\omega;b\omega\bigr) \\
\cB_m & = \prod_{a} \prod_{\ell_a=-j_a}^{j_a} \widetilde{\Gamma}\Bigl( \ell_a\alpha(p/2)- \ell_a\alpha(m-q/2)\omega+\omega_a(\xi)+r_a\nu_R;a\omega,b\omega \Bigr) \;,
\eea
such that $\cZ_m = \cA_m \, \cB_m$. Then $\cZ_\text{tot}$ evaluated on $\hat u$ can be expressed as
\be
\label{eq:ZtotZpmq}
\cZ_\text{tot} \Bigl( \frac{p+q\omega}2; \xi,\nu_R,a\omega,b\omega \Bigr) = \sum_{m=1 \,:\, m'=m}^{ab} \cZ_m + \sum_{(m,m') \,:\, m' \neq m} \bigl( \cZ_m + \cZ_{m'} \bigr) \;.
\ee
Our goal is to show that $\cZ_{m'} = - \cZ_m$.

We begin by considering the contribution of $\cA_m$. Using (\ref{property map gamma}) we can write
\bea
\label{steps A_m'}
\cA_{m'} &= \theta_0 \bigl( p + (2m-q)\omega - 2ra\omega; a\omega\bigr) \, \theta_0\bigl( -p+ (2m-q)\omega + 2sb\omega; b\omega \bigr) \\
&= \theta_0 \bigl( -p - (2m-q)\omega + (2r+1)a\omega; a\omega \bigr) \, \theta_0\bigl( -p+ (2m-q)\omega + 2sb\omega; b\omega \bigr) \;.
\eea
In the second equality we used the second relation in (\ref{eq:theta_0periodicities}). Using the first relation in (\ref{eq:theta_0periodicities}), the identity $2m-q-ra+sb=0$ and reinstating $\alpha$, with some algebra we obtain
\be
\label{eq:Am'Am}
\cA_{m'} = - \, e^{-2\pi i \, \alpha(r) \, \alpha(s) \, \nu_R} \; \cA_m \;.
\ee
Then we turn to $\cB_m$ and, using (\ref{property map gamma}), write
\bea
\cB_{m'} &= \prod_a \prod_{\ell_a=-j_a}^{j_a} \wt\Gamma\Bigl( \ell_a p + \ell_a(2m-q)\omega + \omega_a(\xi) + r_a \nu_R - 2\ell_a r a \omega; a\omega, b\omega \Bigr) \\
&= \prod_a \prod_{\ell_a=-j_a}^{j_a} \wt\Gamma\Bigl( \ell_a p - \ell_a(2m-q)\omega + \omega_a(\xi) + r_a \nu_R + 2\ell_a r a \omega; a\omega, b\omega \Bigr) \;.
\eea
We recall that $j_a$ can be integer or half-integer. In the second equality we simply redefined $\ell_a \to -\ell_a$ and shifted the argument by the integer $2\ell_a p$. Using the identity (\ref{eq:Gammaperiodicities}) repeatedly and distinguishing the cases $\ell_a \lessgtr 0$, we obtain
\be
\cB_{m'} = \Theta \times \cB_m
\ee
where the factor $\Theta$ equals
\be
\Theta = \prod_a \prod_{\ell_a>0}^{j_a} \prod_{k=0}^{2\ell_a r-1} \frac{ \theta_0\bigl( \ell_a p - \ell_a(2m-q)\omega + \omega_a(\xi) + r_a \nu_R + ka\omega; b\omega \bigr) }{ \theta_0\bigl( - \ell_a p + \ell_a(2m-q)\omega + \omega_a(\xi) + r_a \nu_R + (k-2\ell_a r) a\omega; b\omega \bigr) } \;.
\ee
The second product starts from $1$ or $\frac12$ depending on $j_a$ being integer or half-integer.
Using $2m-q-ra+sb=0$ at denominator and shifting the arguments by integers, we rewrite
\be
\Theta = \prod_a \prod_{\ell_a>0}^{j_a} \prod_{k=0}^{2\ell_a r-1} \frac{ \theta_0\bigl( \ell_a p - \ell_a(2m-q)\omega + \omega_a(\xi) + r_a \nu_R + ka\omega; b\omega \bigr) }{ \theta_0\bigl( \ell_a p - \ell_a(2m-q)\omega + \omega_a(\xi) + r_a \nu_R + ka\omega - 2 \ell_a sb \omega; b\omega \bigr) } \;.
\ee
Finally we use the first relation in (\ref{eq:theta_0periodicities}) at denominator, to obtain
\be
\Theta = \prod_a \prod_{\ell_a>0}^{j_a} (-1)^{4\ell_a^2 r s + 8\ell_a^3 r s p} \, e^{-8\pi i \ell_a^2 r s \, \omega_a(\xi)} \, e^{-8\pi i \ell_a^2 r s (r_a-1) \nu_R} \;.
\ee
Reinstating the root $\alpha$, this factor can be written as
\be
\Theta = \prod_{a} \prod_{\ell_a>0}^{j_a} (-1)^{\ell_a^3 \alpha(r) \alpha(s) \alpha(p)} \times \prod_{a,\, \rho_a\in\fR_a} e^{\pi i \rho_a(r) \rho_a(s) \left( \frac12 - \omega_a(\xi) - (r_a-1) \nu_R \right)} \;.
\ee
Combining with (\ref{eq:Am'Am}), the factor picked up by $\cZ$ can be expressed in terms of the anomaly coefficients (\ref{eq:anomalycoeffs}) and (\ref{eq:pertRanomalycoeffs}):
\be
\cZ_{m'} = - \, e^{2\pi i \phi} \; e^{\pi i rs \left( \frac12 \cA^{ii} - \cA^{ii\alpha} \xi_\alpha - \cA^{iiR} \nu_R \right)} \; \cZ_m \;.
\ee
Here $i$ is the gauge index taking a single value. We recall the anomaly cancelation conditions $\cA^{ii\alpha} = \cA^{iiR} = 0$ and $\cA^{ii} \in 4\bZ$, implying that the second exponential equals $1$. In the first exponential we defined
\be
\label{def phi}
\phi = \frac12 \, \alpha(r) \, \alpha(s)\, \alpha(p) \sum_a \sum_{\ell_a>0}^{j_a} \ell_a^3 = 4rsp \sum_a \sum_{\ell_a>0}^{j_a} \ell_a^3 \;.
\ee
It remains to show that $\phi \in \bZ$, so that also the first exponential equals $1$.

For each chiral multiplet in the theory, indicized by $a$, in order to evaluate the second sum in (\ref{def phi}) we should distinguish different cases:
\be
\psi_j \,\equiv\, 4\sum_{\ell>0}^j \ell^3 = \begin{cases}
j^2(j+1)^2 & \in 4\bZ \qquad\;\;\, \text{if } j \in \bZ \\
2(k+1)^2(8k^2+16k+7) & \in 2\bZ \qquad\;\;\: \text{if } j = 2k+\tfrac32 \in 2\bZ + \tfrac32 \\
\frac12 \, (2k+1)^2(8k^2+8k+1) & \in 4\bZ + \tfrac12 \quad \text{if } j = 2k+\tfrac12 \in 2\bZ + \tfrac12 \;.
\end{cases}
\ee
Therefore, chiral multiplets whose gauge representation has spin $j \in \bZ$ or $j \in 2\bZ+\frac32$ give integer contribution to $\phi$. On the other hand, chiral multiplets with $j \in 2\bZ+ \frac12$ can give half-integer contribution.
However, because of the Witten anomaly \cite{Witten:1982fp}, the total number of such multiplets must be even. This is reproduced by the condition (\ref{eq:gaugeanomaly}) on the pseudo-anomaly coefficient $\cA^{ii}$. Indeed, the contribution of a chiral multiplet to the pseudo-anomaly is
\be
\cA^{ii}_{(j)} = \sum_{\ell=-j}^j (2\ell)^2 = \frac43\, j(j+1)(2j+1) \,\in\, \begin{cases} 4\bZ & \text{if $j \in \bZ$ or $j \in 2\bZ + \frac32$} \\ 4\bZ + 2 & \text{if $j \in 2\bZ + \frac12$} \;, \end{cases}
\ee
and the condition $\cA^{ii} \in 4\bZ$ requires that the total number of chiral multiplets with $j\in 2\bZ+\frac12$ be even. This implies that $\phi \in \bZ$, and thus that $\cZ_{m'} = - \cZ_m$. In turn, using (\ref{eq:ZtotZpmq}), this implies that
\be
\cZ_\text{tot}(\hat u; \xi, \nu_R, a\omega, b\omega) = 0
\ee
whenever $\hat u$ is fixed on the torus by the non-trivial element $s_\alpha$ of the Weyl group of $\fsu(2)$.

\subsection{The higher-rank case}
\label{sec:higherrank}

Let us now move to the case of a generic semi-simple gauge algebra $\fg$ of rank $\rk(G)$. The Weyl group $\cW_G$ is a finite group generated by the Weyl reflections
\be
\label{eq:Weylreflection}
s_\alpha(u) = u - 2 \frac{\alpha(u)}{(\alpha,\alpha)} \, \wt\alpha \qquad\qquad \forall \; u \in \fh \;,
\ee
where $\wt\alpha$ is the image of the root $\alpha$ under the isomorphism $\fh^* \to \fh$ induced by the non-degenerate scalar product $( \cdot, \cdot)$ on $\fh^*$. Suppose that there exists a non-trivial element $w$ of $\cW_G$ such that $w \cdot \hat u = \hat u$. It is a standard theorem that the Weyl group acts freely and transitively on the set of Weyl chambers. Therefore, $\hat u$ cannot belong to a Weyl chamber but must instead lie on a boundary between two or more chambers. Such boundaries are the hyperplanes fixed by the Weyl reflections, $\{u | s_\alpha(u) = u\}$, and their intersections. We conclude that there must exist at least one root $\hat\alpha$ such that $s_{\hat\alpha}(\hat u) = \hat u$.

On the other hand, we are interested in points $\hat u$ such that their equivalence class on the torus is fixed by a non-trivial element of the Weyl group, $w \cdot [\hat u] = [\hat u]$. In this case, for each $w$ we can always identify (at least) one root $\hat\alpha$ such that $s_{\hat\alpha} [\hat u] = [\hat u]$, and moreover we can choose a set of simple roots that contains $\hat\alpha$. Let us fix a basis of simple roots $\{\alpha_l\}_{l=1,\dots,\rk(G)}$ for $\fg$ that contains $\hat\alpha$. The fundamental weights $\lambda_l$ are defined by
\be
\label{eq:fundweights}
2 \, \frac{(\lambda_k,\alpha_l)}{(\alpha_l,\alpha_l)}= \delta_{kl} \;.
\ee
We choose a basis $\{H^i\}$ for the Cartan subalgebra $\fh$ such that the fundamental weights have components ${\lambda_l}^i = \lambda_l(H^i) = \delta^i_l$. In this basis $\rho(H^i)\equiv \rho^i\in\nZ$ for any weight $\rho\in\Lambda_\text{weight}$. Moreover, the double periodicity of the gauge variables $u = u_i H^i$ is $u_i\sim u_i + 1 \sim u_i + \omega$. From (\ref{eq:Weylreflection}), the fixed points should satisfy
\be
\label{eq:simplefixedpoint}
2 \frac{\hat\alpha(\hat u)}{(\hat\alpha, \hat\alpha)} \, \widetilde\alpha = p + q \omega \qquad\text{for } \quad p = p_i H^i \;,\quad q = q_i H^i \quad\text{ and }\quad p_i,\,q_i\in\nZ \;.
\ee
Here $\wt\alpha$ is dual to $\hat\alpha$. It is clear that $p,q$ should be aligned with $\wt\alpha$, therefore we set
\be
\label{eq:pandq}
p = \frac{2\hat p}{(\hat\alpha, \hat\alpha)} \widetilde\alpha \;,\qquad\qquad q = \frac{2\hat q}{(\hat\alpha, \hat\alpha)} \widetilde\alpha \;, \qquad\quad \text{ with } \hat p,\,\hat q\in\nZ \;.
\ee
In the basis $\{H^i\}$ we have choosen, the components of $\widetilde\alpha$ are $(\lambda_i, \hat\alpha) = \delta_{il} \, (\hat \alpha, \hat\alpha)/2$, where $l$ is such that $\hat\alpha=\alpha_l$ and we have used \eqref{eq:fundweights}. Only one component of $\widetilde\alpha$ is non-zero, which implies that the integer components of $p,\,q$ are $p_i = \hat p\, \delta_{i l}$ and $q_i = \hat q\, \delta_{i l}$. This proves integrality of $\hat p,\, \hat q$. The general solution to \eqref{eq:simplefixedpoint} can then be written as
\be
\label{eq:simplesolution}
\hat u = \hat u_0 + \frac{p + q \omega}2 \;,
\ee
where $\hat u_0$ is such that $\hat\alpha(\hat u_0) = 0$.

Now, consider the explicit expression (\ref{eq:Ztot}) for $\cZ_\text{tot}$, in terms of $\cZ$ given in (\ref{eq:Z}).
Given any representation $\fR$ of $\fg$, we can always decompose it into irreducible representations of the $\fs\fu(2)_{\hat\alpha}$ subalgebra associated with $\hat\alpha$. The set of weights (with multiplicities) $\Lambda_\fR$ corresponding to $\fR$
can be organized as a union $\Lambda_\fR = \cup_I \Lambda_{\fR,I}$ of subsets $\Lambda_{\fR,I}$, each corresponding to a representation of $\fs\fu(2)_{\hat\alpha}$. Concretely, each $\Lambda_{\fR,I}$ is associated to a representation of $\fs\fu(2)_{\hat\alpha}$ of spin $j_I$, so that its elements can be expressed as an $\hat\alpha$-chain:
\be
\label{eq:lambdaIR}
\Lambda_{\fR,I}=\bigl\{\hat \rho_I+ \ell_I{\hat\alpha} \;\big|\; \ell_I=-j_I,-j_I+1,\dots,j_I-1, j_I\bigr\} \;.
\ee
Here $\hat\rho_I$ is the central point, which is orthogonal to ${\hat\alpha}$, \ie~such that $(\hat \rho_I,{\hat\alpha})=0$. Notice that, in general, $\hat \rho_I$ is not a weight.%
\footnote{Indeed, $\hat\rho_I$ is guaranteed to be a weight (and in particular a root) only if the spin $j_I$ is integer.}
The product over all weights $\rho$ of the representation $\fR$ can then be expressed as a product over the representations of $\fs\fu(2)_{\hat\alpha}$ contained in $\fR$. In particular we can write
\begin{multline}
\label{eq:rhoaGamma}
\prod_{a}\prod_{\rho_a\in\fR_a} \widetilde{\Gamma}\bigl( \rho_a(u)+\omega_a(\xi)+r_a\nu_R;a\omega,b\omega \bigr) = {} \\
{} = \prod_{a, I} \prod_{\ell_{aI} = - j_{aI}}^{j_{aI}} \widetilde{\Gamma}\bigl( \hat \rho_{aI}(u) + \ell_{aI} \hat\alpha(u)+\omega_a(\xi)+r_a\nu_R;a\omega,b\omega \bigr) \;.
\end{multline}
When specifying $\fR$ to the adjoint representation, we obtain a similar decomposition for the roots of $\fg$. Besides the roots $\hat\alpha$ and $-\hat\alpha$ of $\fsu(2)_{\hat\alpha}$, the other roots organize into $\hat\alpha$-chains that we indicate as
\be
\label{eq:lambdaadj}
\Lambda_{\text{roots},J} = \bigl\{ \hat\beta_J + \ell_J \hat \alpha \;\big|\; \ell_J = - j_J , -j_J+1,\dots,j_J-1, j_J \bigr\} \;,
\ee
where $\hat \beta_J$ is the non-vanishing central point orthogonal to $\hat\alpha$ (once again, $\hat\beta_J$ is in general not a weight). Notice that, for each subset $\Lambda_{\text{roots},J}$ of the set of roots, there is a disjoint conjugate subset $\overline\Lambda_{\text{roots},J}$ with the same spin $j_J$ but opposite central point $-\hat\beta_J$.%
\footnote{It is easy to prove that $\Lambda_{\text{roots},J}$ and $\overline\Lambda_{\text{roots},J}$ are disjoint. Suppose, on the contrary, that there exists some common element $\hat \beta_J+\ell_J \hat\alpha = -\hat\beta_J + k_J \hat\alpha$ for some $\ell_J, k_J$. This would imply that $\hat \beta_J = (k_J-l_J)\hat\alpha/2$, but since $(\hat \beta_J,\hat\alpha)=0$, then $\hat\beta_J=0$. Since the only roots proportional to $\hat\alpha$ are $-\hat\alpha$ and $\hat\alpha$ itself, we have reached a contradiction.}
For this reason, we have that
\begin{multline}
\label{eq:ZZ2}
\cZ(u;\xi,\nu_R,a\omega,b\omega) = \theta_0 \bigl( \hat \alpha(u);a\omega \bigr) \, \theta_0\bigl( -\hat \alpha(u);b\omega \bigr) \times {} \\
{} \times\frac{ \prod_{a, I} \prod_{\ell_{aI}=-j_{aI}}^{j_{aI}} \widetilde{\Gamma} \bigl( \hat \rho_{aI}(u)+\ell_{aI}\hat\alpha(u)+\omega_a(\xi)+r_a\nu_R;a\omega,b\omega \bigr) }{ \prod_{J}\prod_{\ell_J=-j_J}^{j_J} \widetilde{\Gamma} \bigl( \hat\beta_J(u)+\ell_J\hat\alpha;a\omega,b\omega \bigr) \, \widetilde{\Gamma} \bigl( -\hat\beta_J(u)+\ell_J\hat\alpha;a\omega,b\omega \bigr)} \;.
\end{multline}


Similarly to the rank one case, we want to prove that $\cZ_\text{tot}(\hat u;\xi,\nu_R,a\omega,b\omega) = 0$ for $\hat u$ in \eqref{eq:simplesolution}. Thus,
we construct an involutive map $\gamma: m \mapsto m'$, acting on the set $\cM$ of vectors $m = m_i H^i$ with integer components $1 \leq m_i \leq ab$. The map is constructed in such a way that it leaves $m$ invariant along the directions orthogonal to $\widetilde\alpha$, whereas it shifts the component parallel to $\widetilde\alpha$ by an integer amount. To be precise, take two vectors $r,\,s\in\fh$ such that 
\be
\label{eq:rands}
r = \frac{2\hat r}{(\hat\alpha, \hat\alpha)} \, \widetilde\alpha \;,\qquad\qquad s = \frac{2\hat s}{(\hat\alpha, \hat\alpha)}\, \widetilde\alpha \;,\qquad\qquad \text{with } \hat r,\, \hat s \in \bZ \;,
\ee
meaning that $r,\,s$ are parallel to $\widetilde\alpha$ and have integer components $r_i = \hat r \, \delta_{il}$, $s_i = \hat s \, \delta_{il}$. Then, we construct $m'$ as
\be
\label{eq:m'=m+sb}
m' = m + s \,b \;,
\ee
which implies that $m'$ differs from $m$ only by integer shifts along the direction of $\widetilde\alpha$.
For $\hat s$ we take the unique integer such that $m' \in \cM$ and
\be
\label{eq:m'malpharelations}
\hat\alpha(m') = \hat\alpha(m) + \hat\alpha(s) \, b = \hat\alpha(q - m) + \hat\alpha(r) \, a \;.
\ee
Indeed, consider the following equation in $r$ and $s$: $2\hat\alpha(m) - \hat\alpha(q) = \hat\alpha(r) \, a - \hat\alpha(s) \, b$. Using \eqref{eq:pandq} and\eqref{eq:rands}, it reduces to $\hat\alpha(m) - \hat q = \hat r a - \hat s b$. Since $a,b$ are coprime, this equation always admits an infinite number of solutions in the pair $(\hat r, \hat s)$, which can be parametrized as $(\hat r_0 + kb, \hat s_0 + ka)$ with $k\in \bZ$. There is however one and only one solution such that $m'$ has components $1\leq m'_i \leq ab$. We define $\gamma(m) = m'$ in such a way. One can easily check that it is an involution.


As in the rank-one case, we adopt the notation
\be
\cZ_m \,\equiv\, \cZ\bigl( \hat u-m\omega;\xi,\nu_R,a\omega,b\omega\bigr) = \cZ \bigl( \hat u_0 +  p/2 - (m-q/2) \omega;\xi,\nu_R,a\omega,b\omega\bigr) \;,
\ee
and, for later convenience, split $\cZ$ into the vector multiplet and chiral multiplet contributions:
\begin{align}
\cA_m &= \theta_0 \bigl( \hat\alpha(p/2) - \hat\alpha (m-q/2)\omega;a\omega\bigr) \, \theta_0\bigl(-\hat\alpha(p/2) + \hat\alpha(m-q/2)\omega; b\omega\bigr) \\
\cC^{\pm}_m &= \prod_J\prod_{\ell_J=-j_J}^{j_J} \widetilde{\Gamma}\bigl( \pm\hat\beta_J(\hat u_0 - m\omega) + \ell_J \hat\alpha(p/2) - \ell_J \hat\alpha(m-q/2) \omega; a\omega, b\omega) \nn \\
\cB_m &= \prod_{a,I} \prod_{\ell_{aI}=-j_{aI}}^{j_{aI}} \widetilde{\Gamma}\Bigl( \hat \rho_{aI}(\hat u_0 - m \omega) +  \ell_{aI} \hat\alpha(p/2)- \ell_{aI} \hat\alpha(m-q/2) \omega + \omega_a(\xi) + r_a\nu_R; a\omega, b\omega \Bigr) \nn
\end{align}
such that $\cZ_m = \cA_m \, \cB_m / \cC^+_m \, \cC^-_m$. We will prove that $\cZ_{m'} = -\cZ_m$, which implies that $\cZ_\text{tot}(\hat u)$ vanishes because $\gamma$ is an involution.

%
%

We begin by considering the contribution of $\cA_{m'}$. Following the same steps as in \eqref{steps A_m'} and using \eqref{eq:m'malpharelations} and (\ref{eq:theta_0periodicities}), we can show
\be
\label{eq:Apm'qpmq}
\cA_{m'} = - e^{-2\pi i \, \hat\alpha(r) \, \hat\alpha(s) \, \nu_R} \; \cA_m \;.
\ee
We also used that $\hat\alpha(r),\, \hat\alpha(s),\, \hat\alpha(q) \in 2\nZ$, which is guaranteed by \eqref{eq:pandq} and \eqref{eq:rands}.
%
%
We now turn to $\cB_{m'}$. Eqn.~\eqref{eq:m'=m+sb} implies that $\hat \rho_{aI}(m') = \hat \rho_{aI}(m)$ for any $\hat \rho_{aI}$ orthogonal to $\hat\alpha$. Using the identity \eqref{eq:Gammaperiodicities} repeatedly and distinguishing the cases $\ell_{aI} \lessgtr 0$, we obtain
\be
\cB_{m'} = \prod_{a,I} \prod_{\ell_{aI}>0}^{j_{aI}} (-1)^{\ell_{aI}^3 \hat\alpha(r) \hat\alpha(s) \hat\alpha(p)} \prod_{a,\, \rho_a} (-1)^{\frac12\rho_a(r)\rho_a(s)} \, e^{-\pi i \rho_a(r) \rho_a(s) \bigl(\rho_a(\hat u_0 - m\omega) +\omega_a(\xi) + (r_a-1) \nu_R \bigr)} \, \cB_m \;.
\ee
The analysis of $\cC^\pm_m$ is analogous to the one for $\cB_m$ and it gives the following:
\be
\label{eq:Cm'm}
\cC_{m'}^\pm = \prod_J \prod_{\ell_J>0}^{j_J} (-1)^{\ell_J^3 \hat\alpha(r) \hat\alpha(s) \hat\alpha(p)} \prod_{\alpha\neq\pm\hat\alpha} (-1)^{\frac12\alpha(r)\alpha(s)} \; e^{\pi i \alpha(r)\alpha(s)\nu_R} \times \cC^\pm_m \;.
\ee
Combining \eqref{eq:Apm'qpmq} with the latter, we obtain that the vector-multiplet contribution is
\be
\cA_{m'} / \cC_{m'}^+ \cC_{m'}^- = 
- e^{-2\pi i \sum_{\alpha\in\Delta}\alpha(r)\alpha(s)\nu_R} \; \cA_m / \cC_m^+ \cC_m^- \;.
\ee
We used $\hat\alpha(r) \hat\alpha(s) \in 4\bZ$, as well as $\sum_{\alpha\in\Delta} \alpha(r) \alpha(s)\in4\nZ$ for any semi-simple Lie algebra $\fg$, and that 
$2\ell_J^3 \hat\alpha(r) \hat\alpha(s) \hat\alpha (p)\in2\nZ$ for any integer or half-integer spin. Including now also the contribution from $\cB_m$, the factor picked up by $\cZ$ can be expressed in terms of the anomaly coefficients \eqref{eq:anomalycoeffs} and \eqref{eq:pertRanomalycoeffs}:
\be
\cZ_{m'} = - \, e^{2\pi i \phi} \; e^{\pi i\, r_i s_j \left( \frac12 \cA^{ij} - \cA^{ijk}(\hat u_0 - m \omega)_k - \cA^{ij \alpha} \xi_\alpha - \cA^{ijR} \nu_R \right)} \; \cZ_m \;.
\ee

The anomaly cancelation conditions $\cA^{ijk} = \cA^{ij\alpha} = \cA^{ijR} = 0$ and $\cA^{ij} \in 4\nZ$ imply that the second exponential equals $1$. In the first exponential we defined
\be
\phi = \frac12 \, \hat\alpha(r) \, \hat\alpha(s)\, \hat\alpha(p) \sum_{a,I} \sum_{\ell_{aI}>0}^{j_{aI}} \ell_{aI}^3 = 4\hat r \hat s \hat p \sum_{a,I} \sum_{\ell_{aI}>0}^{j_{aI}} \ell_{aI}^3 \;.
\ee
Once again, in an anomaly-free theory $\phi \in \bZ$. Indeed, labelling the chiral multiplets by $a$, their $\fsu(2)_{\hat\alpha}$ representations by $I$ and dubbing their spin $j_{aI}$, the only non-integer contributions to $\phi$ come from representations with $j_{aI} \in 2\bZ + \frac12$. On the other hand, the contribution of an $\fsu(2)_{\hat\alpha}$ representation to the pseudo-anomaly coefficient is
\be
\cA^{ij}_{a I} = \sum_{\ell = -j_{aI}}^{j_{aI}} (\hat \rho_{aI} + \ell \hat\alpha)^i(\hat \rho_{aI} + \ell \hat\alpha)^j \;.
\ee
Since generic vectors $r,s$ \eqref{eq:rands} have integer components, the condition $\cA^{ij} \in 4\bZ$ implies that also $\cA^{ij} r_i s_j \in 4\bZ$ for any choice of $r,s$. Contracting with the vectors, we obtain
\be
\cA^{ij}_{a I} r_i s_j = \frac 43 \hat r \hat s \, j_{aI}(j_{aI} + 1)(2j_{aI} + 1) \,\in\, \begin{cases} 4\bZ & \text{if $j_{aI} \in \bZ$ or $j_{aI} \in 2\bZ + \frac32$} \\ 4\bZ + 2 & \text{if $j_{aI} \in 2\bZ + \frac12$} \;. \end{cases}
\ee
Therefore, the condition $\cA^{ij} \in 4\bZ$ requires that the number of $\fsu(2)_{\hat\alpha}$ representations with $j_{aI} \in 2\bZ + \frac12$ be even, and this guarantees that $\phi = 0$.

\bibliographystyle{ytphys}
\bibliography{BHentropy}
\end{document}